\documentclass[acmtog]{acmart}

\copyrightyear{2025}
\acmYear{2025}
\setcopyright{rightsretained}
\setcctype{by}  %

\acmConference[SIGGRAPH Conference Papers '25]{Special Interest Group on

Computer Graphics and Interactive Techniques Conference Conference Papers
}{August 10--14, 2025}{Vancouver, BC, Canada}
\acmBooktitle{Special Interest Group on Computer Graphics and Interactive
Techniques Conference Conference Papers (SIGGRAPH Conference Papers '25),
August 10--14, 2025, Vancouver, BC, Canada}
\acmDOI{10.1145/3721238.3730759}
\acmISBN{979-8-4007-1540-2/2025/08}

\makeatletter
\def\@copyrightpermission{
  \begin{minipage}{\columnwidth}
    
    \href{https://creativecommons.org/licenses/by/4.0/}{\includegraphics[height=2\baselineskip]{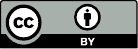}}\\
    {\small\href{https://creativecommons.org/licenses/by/4.0/}{\footnotesize This work is licensed under a Creative Commons Attribution~4.0 International License.}}\\
  
  \end{minipage}%
  \vspace{5pt}%
}
\makeatother

\usepackage{natbib}
\usepackage{algorithm}
\usepackage{algpseudocode}
\usepackage{wrapfig}
\usepackage{subcaption}

\usepackage[capitalize,noabbrev]{cleveref}

\usepackage{siunitx}
\setcitestyle{square}

\newcommand{\param}{w}
\newcommand{\params}{\mathbf{\param}}
\newcommand{\thickness}{t}
\newcommand{\tpms}{f}
\newcommand{\tpmsInterp}{F}

\usepackage{acro}
\acsetup{single}
\DeclareAcronym{FEM}{
    short=FEM,
    long=finite element method
}
\DeclareAcronym{FEA}{
    short=FEA,
    long=finite element analysis
}
\DeclareAcronym{TPMS}{
    short=TPMS,
    long=triply periodic minimal surface
}
\DeclareAcronym{SEM}{
    short=SEM,
    long=scanning electron microscopy
}
\DeclareAcronym{MLP}{
    short=MLP,
    long=multilayer perceptron
}
\DeclareAcronym{MSE}{
    short=MSE,
    long=mean squared error
}
\DeclareAcronym{PCA}{
    short=PCA,
    long=principal component analysis
}
\DeclareAcronym{BO}{
    short=BO,
    long=Bayesian optimization
}

\newcommand{\paperTitle}{Data-Efficient Discovery of Hyperelastic TPMS Metamaterials with Extreme Energy Dissipation}

\usepackage{pdfpages}

\acmDOI{10.1145/3721238.3730759}

\begin{document}

\title{\paperTitle}

\author[M. Perroni-Scharf]{Maxine Perroni-Scharf}
\affiliation{%
  \institution{Massachusetts Institute of Technology}
  \country{USA}
}
\email{max1@mit.edu}
\authornote{Joint first authors with equal contributions.}
\author[Z. Ferguson]{Zachary Ferguson}
\affiliation{%
  \institution{Massachusetts Institute of Technology and CLO Virtual Fashion}
  \country{USA}
}
\email{zy.fergus@gmail.com}
\authornotemark[1]
\author[T. Butruille]{Thomas Butruille}
\affiliation{%
  \institution{Massachusetts Institute of Technology}
  \country{USA}
}
\email{thomasb3@mit.edu}
\author[C. M. Portela]{Carlos M. Portela}
\affiliation{%
  \institution{Massachusetts Institute of Technology}
  \country{USA}
}
\email{cportela@mit.edu}
\author[M. Konaković Luković]{Mina Konaković Luković}
\affiliation{%
  \institution{Massachusetts Institute of Technology}
  \country{USA}
}
\email{minakl@mit.edu}

\begin{abstract}
Triply periodic minimal surfaces (TPMS) are a class of metamaterials with a variety of applications and well-known primitive morphologies. We present a new method for discovering novel microscale TPMS structures with exceptional energy-dissipation capabilities, achieving double the energy absorption of the best existing TPMS primitive structure.  Our approach employs a parametric representation, allowing seamless interpolation between structures and representing a rich TPMS design space. As simulations are intractable for efficiently optimizing microscale hyperelastic structures, we propose a sample-efficient computational strategy for rapid discovery with limited empirical data from 3D-printed and tested samples that ensures high-fidelity results. We achieve this by leveraging a predictive uncertainty-aware Deep Ensembles model to identify which structures to fabricate and test next. We iteratively refine our model through batch Bayesian optimization, selecting structures for fabrication that maximize exploration of the performance space and exploitation of our energy-dissipation objective. Using our method, we produce the first open-source dataset of hyperelastic microscale TPMS structures, including a set of novel structures that demonstrate extreme energy dissipation capabilities, and show several potential applications of these structures.
\end{abstract}



\begin{teaserfigure}
  \centering
  \vspace*{0.06in}
  \includegraphics[width=0.89\textwidth]{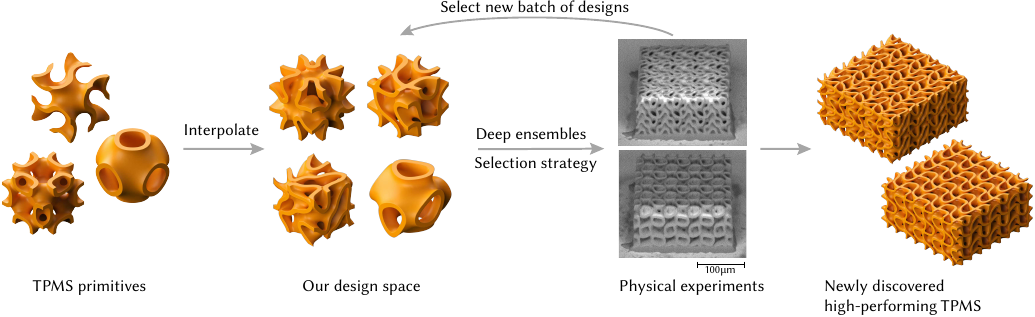}
  \vspace*{-0.105in}
  \caption{Starting from a small set of well-studied TPMS metamaterial primitives, we use their implicit representations and interpolate between them to obtain a rich design space of TPMS structures exhibiting a wide variation in mechanical performance. To model the complex space of their behaviors, we physically test their performance and train Deep Ensembles that capture the uncertainty in prediction and use the results to inform our selection strategy for which designs to fabricate and test next. We iteratively select batches of designs for 3D-printing and testing to improve the model and discover new TPMS structures with higher energy dissipation than in known primitive structures.}
  \Description{}
  \label{fig:teaser}
  \vspace*{0.04in}
\end{teaserfigure}

\maketitle

\smallskip
\section{Introduction}

Metamaterials are artificially engineered materials composed of three-dimensional microstructures, designed to exhibit properties that cannot be achieved by monolithic materials alone \cite{jiao2023mechanical, Xia2022}. These materials can achieve unique electromagnetic \cite{walser2001electromagnetic}, acoustic \cite{Matlack2016}, or mechanical characteristics \cite{Bauer2017}, unlocking innovative applications in a variety of domains. The study of metamaterials is an increasingly active area of research, significantly aided by advances in computer graphics and machine learning, which have enabled new possibilities for optimizing the design of 3D-printable metamaterial structures that exhibit unique or extreme physical properties. These properties include materials with negative Poisson ratios (auxetic materials) \cite{Shim2013}, negative compressibility \cite{gatt2008negative, nicolaou2012mechanical}, ultra-high stiffness \cite{Crook2020,Zheng2014ultralight}, ultra-light weight \cite{schaedler2011ultralight}, and programmable properties and mechanisms \cite{Xia2022}.

Identifying microstructures that can produce such properties poses a significant inverse design challenge. To address this challenge, recent research has increasingly relied on deep neural networks to automate the optimization of microstructures for desired physical responses~\cite{li2023neural, zhang2022uncertainty}. A limitation of most of these approaches is that they rely on simulated data for model training---as generating sufficient physical data through 3D printing and testing is prohibitively expensive and time-consuming---and thus offer no guarantee that the resulting models accurately mirror real-world behaviors. As we show in our work, this simulation-to-reality gap is especially prevalent in settings where fabricated structures are very sensitive to the types of constituent materials and conditions in which they are tested. The limited number of studies that incorporate real-world data into training processes typically do not address strategies for efficient data acquisition through 3D-printing.

Our study addresses these challenges by applying a Bayesian framework to optimize hyperelastic \ac{TPMS} metamaterials. Bayesian optimization iteratively suggests experiments to evaluate that are likely to yield the greatest performance improvement. Our optimization method is uncertainty-aware and based on real-world experimental data. We use Deep Ensembles (DEs)~\cite{DE2017} to capture the uncertainty in predictions and model the complex space of TPMS behaviors. DEs are iteratively refined through a limited number of real-world experiments. The uncertainty measures allow us to target new regions of the parameter and performance spaces as we print subsequent batches of data. Furthermore, we exploit a specific objective for discovering structures with a high capacity to dissipate energy.

To obtain physical experimental data, we utilize sub-micron resolution 3D printing technology---termed two-photon lithography---to fabricate microscopic metamaterials, of approximately 200$\times$200$\times$100 \textmu{}m$^3$ in size. Fabricating samples through this technique enables high-throughput prototyping and subsequent characterization, while ensuring mechanical properties of the constituent polymer are not scale-dependent (i.e., no size effects are present). Given the widespread use of \ac{TPMS} architectures in impact absorption applications \cite{zhang2023study, zhao2024compressive}, our goal is to discover structures with exceptional energy absorption capabilities when compressed by up to 60\% of their original height. \ac{TPMS} structures inherently possess a high surface area-to-volume ratio, are lightweight, and distribute stress more homogeneously than classical truss-based metamaterials. We choose to focus on large-deformation mechanical responses because the combination of high energy dissipation with lightweight properties can result in robust materials suitable for various applications, including bone-scaffolding and impact absorption.

 \begin{figure*}
   \centering
   \vspace*{-0.01in}
   \includegraphics[width=0.915\linewidth]{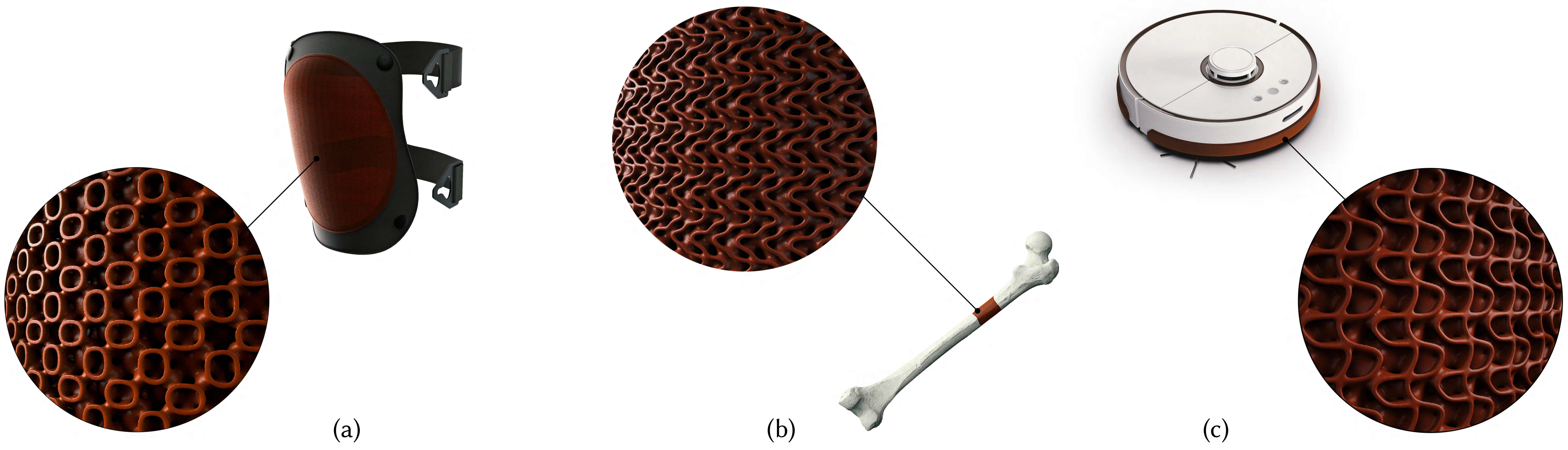}
   \vspace*{-0.07in}
   \caption{Three potential applications for our discovered hyperelastic \acs{TPMS} structures: (a) lightweight, thin, flexible and breathable knee padding, (b) a porous bone implant allowing for cell growth and infiltration, easy to print for patient-specific constraints, and (c) a thin bumper around a robotic vacuum cleaner, providing lightweight protection.}
   \Description{}
   \label{fig:applications}
   \vspace*{-0.025in}
 \end{figure*}

\Cref{fig:applications} illustrates three potential uses for our hyperelastic \ac{TPMS} metamaterial structures. One is lightweight, breathable knee padding, offering flexibility and reducing heat buildup, suitable for wearables and sports equipment. Another is patient-specific bone scaffolds, which could conform to complex anatomical shapes, support cell growth, and allow nutrient transport, building on prior studies \cite{dong2021application}. Lastly, these structures could serve as impact-absorbing bumpers for robot vacuum cleaners, providing lightweight protection without compromising efficiency.

Our findings suggest broader implications for design workflows. While simulation-based pipelines are prevalent in metamaterial discovery~\cite{li2023neural,huang2023optimized}, they often struggle to capture dissipation-related behaviors driven by contact, buckling, and other nonlinear effects at the microscale. Recent work has addressed these challenges via differentiable simulators~\cite{Huang2024Differentiable} or physics-informed models~\cite{peng2023machine}, but these approaches remain limited in capturing emergent behaviors such as jamming or folding. By integrating real-world testing into the optimization loop, our method complements such tools and offers a general strategy for discovering structures whose performance cannot be reliably predicted from simulation alone.

Our key contributions are:
\vspace*{-0.04in}
\begin{enumerate}
    \item The first method that utilizes Bayesian optimization to discover extremal hyperelastic \ac{TPMS} metamaterial structures.
    \item The first physically validated open-source microscale \ac{TPMS} metamaterial dataset, which includes a set of novel structures with extreme energy dissipation capabilities.
    \item A study on the application of traditional finite element method (FEM) simulations to microscale hyperelastic structures.
\end{enumerate}

\Cref{fig:teaser} provides an overview of our method. We first discuss our \ac{TPMS} design representation in Section~\ref{sec:designsp}. Next, we detail our experimental fabrication and testing setup in Section~\ref{sec:experimental}. We then address the limitations of simulations in matching experimental data and optimizing microscale \ac{TPMS} structures in Section~\ref{sec:simulation}. Finally, in Section~\ref{sec:optimization}, we present our primary contribution: an optimization method for discovering novel \ac{TPMS} structures with exceptional energy absorption capabilities.

\section{Related Works}

\paragraph{Metamaterial Design}
Metamaterials have been found to have mechanical properties unprecedented in bulk material systems. Early work on metamaterials by \citet{Deshpande2001} and \citet{Ashby2006} involving stochastic and periodic beam-based lattices explored the physical reasons supporting their ultra-high stiffness-to-weight ratios, launching broad explorations into how metamaterial design parameters, including relative density, constituent material, and topology, might affect mechanical performance, as categorized by \citet{Bauer2017} and \citet{Zhang2020} among others. Beyond exceptional stiffness up to the theoretical boundary for porous materials \cite{Crook2020, Wang2022}, metamaterial designs have achieved extraordinary mechanical properties such as negative Poisson's ratios \cite{Babaee2013}, enhanced fracture toughness \cite{shaikeea2022}, extreme compliance and resilience at large-strain deformation \cite{Surjadi2024}, and finally optimized energy dissipation through the formation of compaction zones \cite{Hawreliak2016, Lind2019, Dattelbaum2020}.

Optimizing energy dissipation in metamaterials for impact mitigation is of special interest due to the wide range of engineering applications where dynamic loading ultimately leads to failure. \citet{Deshpande2000} explored how the formation of compaction shocks in stochastic foams enhanced energy dissipation and how the stress state in the compaction shocks was related to the quasistatic behavior of these foams. Experimental studies on the dynamic response in stochastic \cite{Deshpande2000, Barnes2014}, periodic beam-based \cite{Mines2013, Weeks2023} and shell-based \cite{Tancogne-Dejean2019, Novak2023} metamaterials, including TPMS structures, have demonstrated the viability of these material design frameworks for the purpose of energy dissipation. However, most of these studies have achieved these results without fully taking advantage of advanced quantitative tools such as machine learning and neural architectures for optimizing metamaterial topologies inside these frameworks.

Several key studies in metamaterial design have made significant contributions without machine learning, including the seminal work in computer graphics of  \citet{Bickel2010} introducing computational methods for designing custom materials with specific mechanical properties. \citet{Schumacher2015Microstructures} developed a framework for creating microstructures that can be optimized to change properties across the given shape. \citet{panetta2015elastic} explored compliant mechanisms using metamaterials. \citet{Tzoni2020Low} proposed a parametric representation for flexible planar metamaterials. \citet{Martinez2016,Martinez2018} used Voronoi Diagrams for novel metamaterial designs. \citet{makatura2023procedural} developed a unified procedural graph representation for cellular metamaterials, facilitating the creation of tileable structures with diverse properties. 

Recent advances in metamaterial design have leveraged neural networks to develop encoded implicit parameterizations of metamaterials, primarily trained on simulated data. \citet{lee2022t} used a variational autoencoder to learn latent space representations for 2D pixelated metamaterials. \citet{wang2020deep} and \citet{zhang2022uncertainty} both employed deep generative models to organize microstructures within structured latent spaces. However, such parameterization approaches provide no assurance that decoded metamaterials will be free from structural anomalies like floating islands, thus limiting their applicability for designing 3D validated datasets.

Several studies have tailored neural network architectures to enhance performance predictions for specific structural subclasses. These models typically simulate thousands of structures rapidly, but cannot ensure high fidelity in real-world fabrication scenarios. \citet{bastek2022inverting} inverted the property maps of truss structures based on extensive simulated data. \citet{li2023neural} developed a neural network model to predict the behavior of extruded isohedral tilings, though this model does not account for out-of-plane behaviors or structures with three-dimensional variability. \citet{fang2019deep} proposed deep physically-informed neural networks for the design of electromagnetic metamaterials. \citet{meyer2022graph} investigated the use of graph neural networks to model complex dependencies within metamaterial structures. \citet{oktay2023neuromechanical} introduced neuromechanical autoencoders that integrate a differentiable simulator with neural networks to model intricate two-dimensional geometries. \citet{liu2024programmable} developed programmable neural networks that adapt to various simulation environments.

Our research diverges from these works by integrating data from limited real-world experiments into a robust predictive model, enhancing both the real-world applicability of our designs and discovery efficiency. Unlike the approaches of \citet{ha2023rapid}, who trained their models on a narrow range of strut-based microstructures from a small, 3D-printed dataset, our methodology incorporates a broader experimental base to improve prediction reliability. In alignment with this work, \citet{thakolkaran2023experiment} fabricate several hundred spinodal metamaterials for the training of a predictive model. However, unlike ours, their structures are not strategically selected and the model leverages physics-based inductive biases, which are challenging to define for many other classes of microstructure parameterizations.

\begin{figure}
  \centering
  \includegraphics[width=0.95\linewidth]{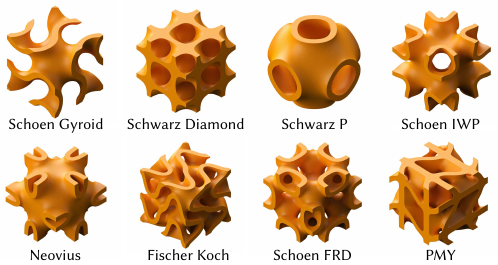}
\vspace*{-0.075in}
  \caption{Unit cells of TPMS primitives used in interpolation to represent the entire design space. Each of these structures shares a periodicity of $2\pi$ in each direction.}
  \Description{}
  \label{fig:primitives}
  \vspace*{-0.15in}
\end{figure}

\vspace*{-0.05in}
\paragraph{Triply Periodic Minimal Surfaces (TPMS)} There are many different subclasses of metamaterials \cite{zadpoor2023design, jiao2023mechanical}, each of which has particular strengths in different applications. Here we focus on \ac{TPMS}~\cite{hu2020efficient}, tileable minimal surfaces that divide space into two continuous regions. TPMS structures cover a broad spectrum of physical properties and applications, from heat transfer~\cite{attarzadeh2022multi, fan2022thermal} to tissue engineering~\cite{feng2021isotropic}, offering unique properties that are beneficial across various domains. We adopt the formulation from \citet{AlKetan2021MSLattice}, representing a vast TPMS metamaterial design space with eight parameters, enabling smooth interpolation between designs. To our knowledge, ours is the first study to efficiently discover novel TPMS metamaterials with extreme energy dissipation and to obtain insights via high-throughput prototyping and characterization of TPMS metamaterials with hyperelastic constitutive properties at the microscale.

\paragraph{Simulation of Metamaterials}
Previous studies rely on simulated data for either building a library of homogenized material properties~\cite{panetta2015elastic,Schumacher2015Microstructures,Tzoni2020Low,huang2023optimized} or training neural networks to predict metamaterial behavior~\cite{li2023neural}. Because of the complexity of simulating full-scale structures, these works often employ \emph{periodic homogenization}~\cite{Chen2021Bistable,Nakshatrala2013Nonlinear}. Furthermore, several studies focus entirely on 2D structures~\cite{li2023neural,huang2023optimized,Tzoni2020Low}, which are easier to simulate and fabricate but have limited real-world applicability. We focus on 3D structures, specifically TPMS, which are more challenging to simulate and fabricate but have a broader range of applications. The use of differentiable simulators for metamaterial design has been explored in several works~\cite{panetta2015elastic,panetta2017worst,huang2023optimized,Li2022Digital,Tzoni2020Low}. These works utilize a variety of optimization techniques but are limited to specific classes of metamaterials and do not explore the design space of TPMS metamaterials.

\paragraph{Guided Data Collection for Material Discovery}
When dealing with experimental design problems for which datasets are not available, collecting new data can be expensive and time-consuming. In this context, \ac{BO}~\cite{jones1998efficient,BOsurvey} has emerged as an effective method for guiding the search for an optimal solution in various applications. \citet{Erps2021} use Bayesian optimization to guide the discovery of better-performing 3D printing resins. \citet{Sharpe2018} explore the design of lattice structures via Bayesian optimization in simulation. \citet{honeycomb2023} optimize the thickness of hexagonal honeycomb metamaterial lattices with Bayesian optimization. \citet{zhang2022uncertainty} provide a comparative study on Bayesian optimization for material design.

Recently, \citet{snapp2024superlative} employed a self-driving lab using Gaussian Processes (GPs) to discover high-performing FDM printed structures. While both studies focus on data-driven optimization of mechanical properties, we address a different domain by modeling hyperelastic microscale structures. Additionally, we utilize a neural network-based surrogate model in place of GPs. This contrast in modeling approaches and targeted applications offers complementary insights into the application of Bayesian optimization in structure design.

Recent advances in the mechanics and materials science communities further highlight the potential of such data-driven strategies. \citet{serles2025ultrahigh} applied Bayesian optimization to identify carbon nanolattices with record-breaking specific strength, leveraging simulations and constrained experimental fabrication. Similarly, \citet{peng2023machine} introduced a constrained multi-objective framework to explore trade-offs between mechanical performance and manufacturability in architected materials. These approaches demonstrate how machine learning and optimization can transform structural material discovery when coupled with domain-specific priors and physical experimentation.

Our work contributes to this growing space by targeting a distinct objective—maximizing energy dissipation in 3D-printed hyperelastic TPMS lattices—under experimental and computational constraints where high-fidelity simulations are infeasible. Unlike the above approaches, which typically use Gaussian Processes, we use Deep Ensembles to represent the surrogate model, enabling uncertainty-aware predictions in complex performance spaces. We also adapt the acquisition function for large batch selection under fabrication budget limits, demonstrating the versatility of Bayesian frameworks across diverse structural material domains.

\section{TPMS Design Space}
\label{sec:designsp}

\begin{figure}
  \centering
  \includegraphics[width=0.84\linewidth]{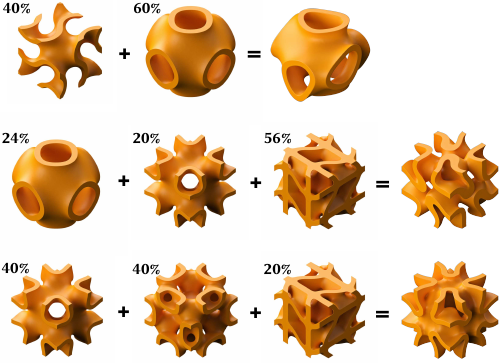}
  \vspace*{-0.07in}
  \caption{Our designs are obtained as a weighted sum of eight TPMS primitives. Each design cell is fully parameterized with a set of weights. For example, the cell in the first row is a combination of 40\% of Schoen Gyroid and 60\% of Schwarz P.}
  \Description{}
  \vspace*{-0.135in}
  \label{fig:interpolation}
\end{figure}

The most popular and well-studied TPMS structures, which we call TPMS primitives, are shown in \cref{fig:primitives}. To represent a wider range of TPMS structures, we model our design space after the one proposed by \citet{AlKetan2021MSLattice}. That is, we take these eight TPMS primitives each expressed as an implicit function (we refer to \citet{AlKetan2021MSLattice} for the definitions) then interpolate between them using barycentric coordinates: \\[-6pt]
\[
\tpmsInterp(x, y, z) := \sum_{i=1}^{8} \param_i \tpms_i(x, y, z),
\] \\[-6pt] 
where $\param_i$ are the weights for each primitive, $\tpms_i$ is the corresponding implicit function, and $\tpmsInterp$ is the resulting interpolated structure's implicit function. Examples of three such interpolations are shown in \cref{fig:interpolation}. This ability to seamlessly interpolate between structures enables potential downstream applications where one could place structures from our dataset in various sections of a larger lattice and smoothly transition between them, creating spatially varying physical properties within a single object \cite{Schumacher2015Microstructures}.

The zero level set of the implicit function represents the \emph{sheet-networks} of the structure. To fabricate these structures, we need to assign a thickness to the sheets. This is commonly done by adding an offset $\pm \tfrac{1}{2}\thickness$ to $\tpmsInterp$~\cite{AlKetan2021MSLattice}:
\[
\left(\tpmsInterp + \tfrac{1}{2}t\right) \left(\tpmsInterp - \tfrac{1}{2}t\right) = 0.
\]

\begin{wrapfigure}[5]{I}[0pt]{0.3\linewidth}
    \begin{center}
        \vspace*{-0.01in}
        \includegraphics[width=0.98\linewidth]{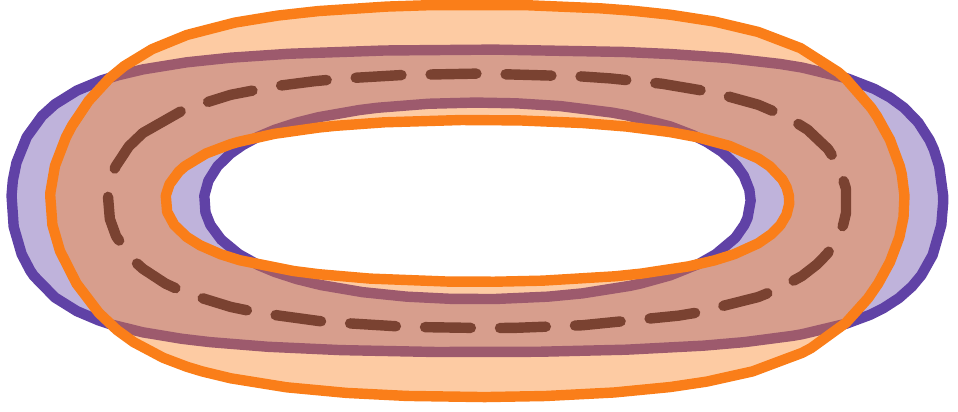}
    \end{center}
    \Description{2D slice of the Neovius primitive using a constant thickness (purple) and our proposed spatially varying thickness (orange).}
    \label{fig:uniform-thickness}
\end{wrapfigure}

This offset, however, is not in standard units of distance and, as we show in purple in the inset figure, results in uneven wall thicknesses. To address this issue, we instead propose a spatially varying thickness and scale the offset by $1/\|\nabla \tpmsInterp\|$ to ensure the walls have a uniform thickness (orange curves):
\[
\left(\tpmsInterp + \tfrac{1}{2}\tfrac{t}{\|\nabla \tpmsInterp\|}\right) \left(\tpmsInterp - \tfrac{1}{2}\tfrac{t}{\|\nabla \tpmsInterp\|}\right) = 0.
\]
We use a thickness of $\thickness=\qty{0.5}{\um}$ in all our experiments.

We mesh these structures using CGAL~\cite{cgal} as a 4\texttimes4\texttimes2 tiling with side lengths of 200$\times$200$\times$100 \textmu{}m\textsuperscript{3}. We choose CGAL over marching cubes because CGAL provides a tetrahedral mesh directly from our implicit function. These tetrahedral meshes are necessary for the \ac{FEA} in \cref{sec:simulation}. Additionally, we use CGAL to produce periodic tetrahedral meshes for periodic simulations. We discard parametric combinations that produce cavities and keep only the largest disconnected component (i.e., discarding isolated material regions).

\section{Experimental Setup}
\label{sec:experimental}

\subsection{Limitations of Simulation-Based Approaches}
\label{sec:simulation}

While prior work has successfully used simulated datasets---par\-ticularly with periodic boundary conditions---to train predictive models~\cite{li2023neural,panetta2015elastic,Schumacher2015Microstructures}, the assumptions in those studies do not fully align with our experimental setup.

We tested both periodic simulations of single unit cells using PolyFEM~\cite{polyfem} and full-scale simulations of 4\texttimes4\texttimes2 structures with rigid plate contact and substrate adhesion. However, calibrating simulation parameters—such as friction coefficients, geometric tolerances, and material properties—is particularly challenging due to inherent uncertainties introduced by our fabrication and development processes, causing a mismatch between simulated and real curves. This is further discussed in the supplementary material.

\vspace*{-0.05in}

\begin{figure}[t]
  \footnotesize
  \includegraphics[width=0.95\linewidth]{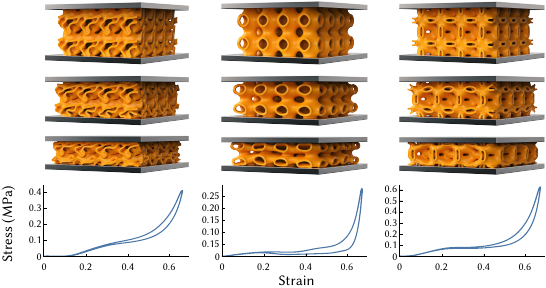}
  \vspace*{-0.09in}
  \caption{Renders of three \ac{TPMS} structures being compressed, along with their experimental stress-strain curves. Rows present 0, 25\%, and 50\% compression of each structure respectively.}
  \Description{}
  \label{fig:fs-sim}
  \vspace*{-0.1in}
\end{figure}

\begin{figure}
  \centering
  \vspace*{0.15in} 
  \includegraphics[width=\linewidth]{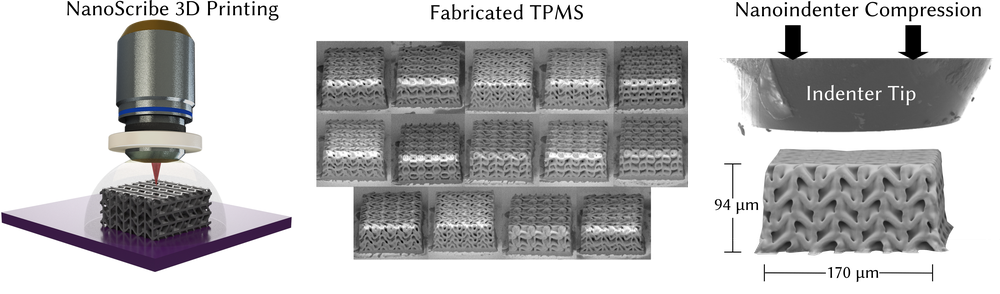}
  \vspace*{-0.25in}
  \caption{Fourteen fabricated micro-TPMS structures printed on a Nanoscribe 3D printer and tested using an Alemnis nanoindenter.}
  \label{fig:fab}
  \Description{}
  \vspace*{-0.15in}
\end{figure}

\subsection{Fabrication}
Benchtop, microscale experiments involving the uniaxial compression of 3D printed TPMS metamaterials are performed to capture sufficient data for Bayesian optimization of their properties, illustrated in \cref{fig:fs-sim}. Here, TPMS metamaterials are first fabricated out of a hyperelastic photosensitive resin (IP-PDMS) using two-photon lithography (NanoScribe Professional GT2), where a femtosecond laser is focused through an optical objective and rastered in three dimensions to produce the TPMS architectures in a layer-by-layer manner, on top of a silicon substrate (Figure~\ref{fig:fab}). To ensure high-fidelity prints, we employ a 25$\times$ magnification objective with a slicing distance (direction normal to the substrate) set to 1.0\,\textmu{}m and a hatching distance of 0.5\,\textmu{}m. The laser speed during printing is set to 13,500 \textmu{}m/s using a laser power of 60\,mW. A polymerization reaction occurs within a microscale ellipsoidal volume (the voxel) where the probability of multi-photon absorption is highest. This fabrication process differs from the more common fused deposition modeling (FDM) 3D printers in its ability to create structures with sub-micron features that lack interfaces between layers (due to voxel overlap between layers) and complex interior features and overhangs without the need for supports. The resulting 3D printed metamaterials are thus composed of an isotropic, hyperelastic polymer that exhibits properties similar to PDMS at the macroscale. It is important to note that polymers printed through two-photon lithography are isotropic and insensitive to print hatching direction, since polymerization occurs stochastically and slicing and hatching distance are chosen so each voxel pass overlaps with former voxel paths. After printing, the TPMS metamaterials are developed and the uncured resin washed away first with a 10-minute wash of isopropyl alcohol followed by a 1-minute rinse in clean isopropyl alcohol.

Each TPMS lattice is a 4$\times$4$\times$2 tessellation of an individual cubic unit cell in $x$, $y$, and $z$ directions respectively. The unit cells have a nominal side length of 50\,\textmu{}m and wall thickness of 4\,\textmu{}m. The as-fabricated dimensions of the TPMS lattices are measured with scanning electron microscopy (SEM) micrographs of gold-coated, uncompressed samples, having an average width and depth of 170$\pm$2 \textmu{}m and an average height of 94$\pm$2 \textmu{}m. These dimensions are used in calculating the stress-strain relations following uniaxial compression for each TPMS lattice sample. Prior to compression, every fabricated TPMS lattice sample is checked for print errors with an SEM to ensure experiments for each lattice are representative of their architectures.

\vspace*{-0.04in}
\subsection{Mechanical Characterization}
Each metamaterial is uniaxially compressed {\it{ex situ}} using a displace\-ment-controlled nanoindenter (Alemnis ASA) with a \qty{400}{\micro\m} diameter flat-punch tip. The compression is performed at a strain rate of 1 s$^{-1}$ to a maximum strain of approximately 55\%. At the beginning of each uniaxial compression experiment, the tip begins displacement \qty{5}{\micro\m} from a TPMS metamaterial sample such that it can accelerate to the desired tip velocity by the time contact is initiated with the sample. Some variation in maximum strain for each sample is expected due to slight variations in sample height and tip distance from sample. During uniaxial compression, displacement is recorded through the piezoelectric actuator that drives the compression, and force is measured through a load cell to which the tip is attached. The force-displacement data---along with the sample dimensions---are then used to calculate the uniaxial engineering stress-strain response for each sample. 

We expect the majority of contributions to the stress response to be from material deformation. The marked increase in the stress response at high strain is caused by contact between walls within the TPMS samples, resulting in a compacted state. Frictional contributions at a state of high strain are difficult to quantify due to their highly nonlinear nature but may slightly hasten the onset of compaction by halting the relative movement between walls. Frictional energy dissipation from the TPMS walls sliding against each other is expected to be minimal in comparison to material deformation.

A minimum of two experiments for each sample architecture are performed to ensure measurement repeatability, with the experiments averaged together and denoised using a Savitzky-Galoy filter \cite{schafer2011savitzky}, from which 120 evenly spaced strain-stress values are selected to pass to the machine learning model for training and model performance verification.
\vspace*{-0.05in}

\begin{figure}
  \centering
  \vspace*{0.075in}
  \includegraphics[width=0.95\linewidth]{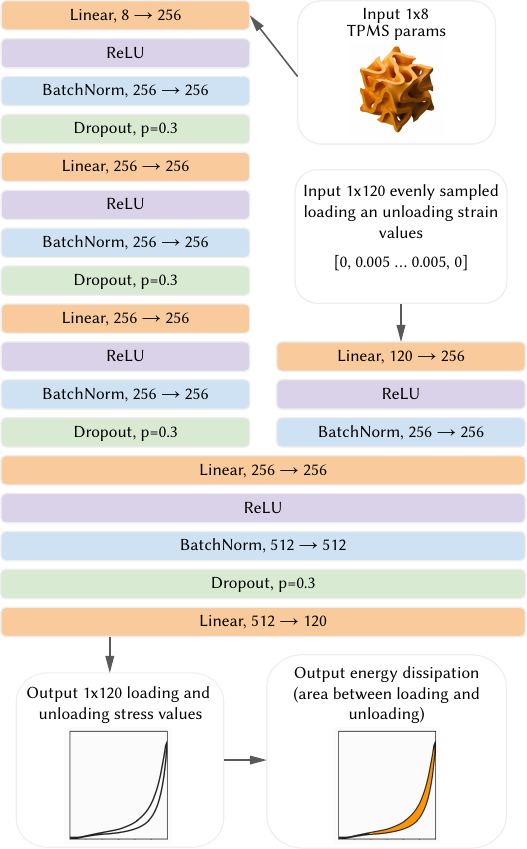}
  \vspace*{-0.075in}
  \caption{Architecture of a single MLP model in the deep ensemble. One head takes in TPMS parameters, the other head takes in strain values. The output is a stress-strain curve, from which we can calculate energy dissipation.}
  \label{fig:modelarch}
  \Description{Single model architecture}
  \vspace*{-0.155in}
\end{figure}

\smallskip
\section{Optimization Framework}
\label{sec:optimization}
\begin{wrapfigure}{I}{0.4\linewidth}
    \centering
    \begin{center}
  \vspace*{-0.06in}
    \includegraphics[width=0.95\linewidth]{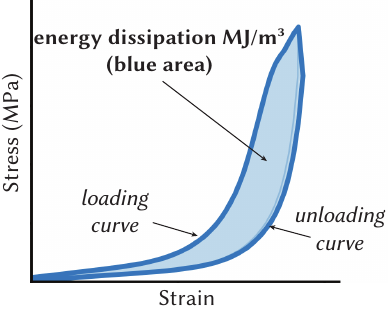}
    \end{center}
    \vspace*{-0.1in}
    \Description{Energy dissipation as calculated by the area between loading and unloading portions of a stress-strain curve.}
    \label{fig:energydiss}
\end{wrapfigure}
Due to the intractability of simulations and the need for high-fidelity results, we use only 3D-printed and tested data. Our goal is to efficiently discover novel TPMS structures with superior energy dissipation, given fixed, limited 3D printing costs. Energy dissipation (\unit{\mega\J\per\m^3}), measured by the area between the loading and unloading portions of a stress-strain curve (see inset), quantifies a structure's energy absorption ability.

To efficiently identify high-performing structures, we avoid the naive approach of uniformly or randomly sampling and testing points in the design space, which would be prohibitively costly. Instead, we propose a data-efficient method using an uncertainty-aware surrogate model to iteratively guide the sampling process. 

We explore the design space by fabricating and testing batches of structures. The surrogate model then helps us decide on what structures to include in the next batch by balancing two objectives: exploring the design space to enhance accuracy and exploiting promising high-energy dissipation regions that we have already identified. This approach allows us to quickly and efficiently identify the best-performing designs, with our best structure exceeding the average uniform sample by more than a factor 20 and achieving twice the energy density of the best primitive TPMS structure.

Below, we outline the steps in our Bayesian optimization-based pipeline, from modeling the performance space to using our model to select the next batch to 3D-print and test.

\vspace*{-0.01in}

\begin{figure}
  \centering
  \includegraphics[width=0.95\linewidth]{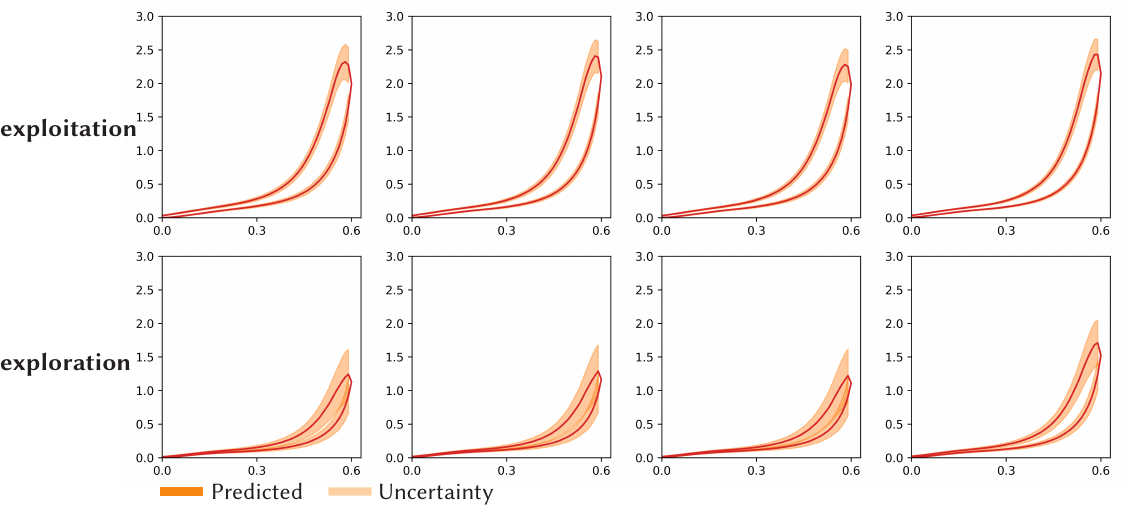}
  \vspace*{-0.05in}
  \caption{Study on the effect of $\kappa$ for batch acquisition. We choose a batch of size four using both $\kappa=0$ and $\kappa=10$. The top row corresponds to the top four scoring samples when $\kappa=0$, whereas the bottom row is the top four selected points when $\kappa=10$. Larger $\kappa$ values encourage exploration by maximizing uncertainty, whereas smaller $\kappa$ values encourage exploitation by maximizing the energy dissipation of our structures.}
  \label{fig:kappa}
  \Description{}
  \vspace*{-0.12in}
\end{figure}

\begin{figure*}
  \centering
  \includegraphics[width=\linewidth]{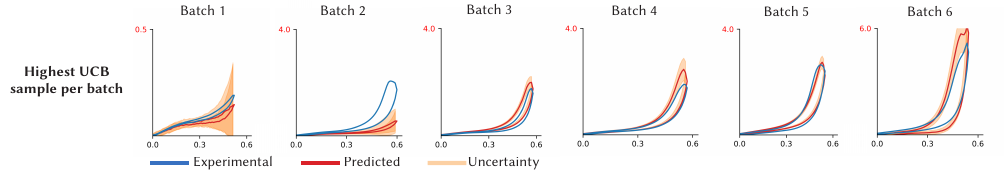}
  \vspace*{-0.25in}
  \caption{We present the experimental (blue), predicted (red), and uncertainty (light orange) stress-strain curves for the highest scoring $\operatorname{UCB}$ sample in successive batches. This sample represents the optimal choice suggested by our optimization algorithm for each batch. The predicted curves are generated by a surrogate model trained on all data up to but excluding the current batch. It is important to note that the $y$-axis scale of the plots increases to accommodate the rise in maximum stress as optimization progresses. Initially, our algorithm selects points with high uncertainty to encourage exploration, transitioning to more exploitative and optimistic points in the later stages.}
  \label{fig:network}
  \Description{}
  \vspace*{-0.03in}
\end{figure*}

\subsection{Neural Network Surrogate Model}

\paragraph{Architecture} To capture the complex range of behaviors exhibited by our TPMS structures, we use Deep Ensembles~\citep{DE2017} as a surrogate to predict the structure behavior based on their parameters, as prior research has demonstrated that ensembles outperform Gaussian Processes in terms of convergence time to optimal solutions and handling complex, non-convex performance spaces \cite{lim2021extrapolative, tian2024boundary}. Our deep ensemble comprises 30 independently trained neural networks, enabling us to obtain both mean predictions and measures of prediction uncertainty that can inform our sampling strategy and improve accuracy.

A key challenge in Bayesian optimization is finalizing the model architecture without access to a large dataset, as collecting extensive experimental data upfront wastes fabrication resources and time. To address this, we used a proven two-branch architecture for predicting stress-strain curves in metamaterials \cite{li2023neural}. Preliminary tests on simulated structures and initial experimental data informed dropout values and hidden layer sizes, reducing overfitting on small datasets.

Each model in our ensemble is a dual-headed MLP (details in \cref{fig:modelarch}) adapted from the neural architecture in \cite{li2023neural} to predict stress-strain behavior. One head processes the eight TPMS parameters, while the other handles an array of evenly applied strain values ranging from 0 to the maximum applied strain for each structure. Treating strain values as a model's input rather than a constant is necessary due to slight variations in the maximum strain applied during testing. This ensures unbiased exploration when sampling for our next batch, preventing undue favoritism toward structures subjected to slightly extra strain.

As we are exploring the uncharted territory of the performance of microscale hyperelastic TPMS structures, we lack data to design a more complex model architecture that performs well with our specific type of structure. Therefore, we keep our model as agnostic as possible and exclude any inductive biases that might limit our model's ability to explore and interpret this new performance space.

We first train our model on an initial batch of 23 structures. Given the very small size of this initial training set, we take careful steps to mitigate overfitting, which would hinder the algorithm's ability to explore new regions of the performance space. To address this, we add dropout \cite{srivastava2014dropout} and batch normalization layers \cite{santurkar2018does} to the model and implement early stopping at convergence during training \cite{caruana2000overfitting}. 

\vspace*{-0.04in}
\paragraph{Training} During each phase of the optimization process, we train each network in our ensemble model using all available physically validated data, including input TPMS parameters and the full range of applied strain values, paired with the corresponding stress values obtained from experimental testing. In the inference phase, the model receives specific metamaterial parameters along with a predefined array of strain values, uniformly sampled from 0 to the maximum applied strain (approximately 0.6). Throughout the training phase, we use the Adam optimizer and the \ac{MSE} loss function between predicted and real stress values. We use a learning rate of 0.001 and train each single \ac{MLP} for up to 2,000 epochs with early stopping patience of 100. 

\vspace*{-0.03in}
\paragraph{Inference} To obtain predicted stress-strain curves, each model $M_i$ in the ensemble is used to predict an array of stress values $S$ from input strain array $E$ and structure parameters $w$. Our final prediction is then the mean of stress-strain predictions across all the $N$ models in the ensemble, $\mu_S(w, E) = \frac{1}{N}\sum^N_iM_i(w, E)$. To quantify the uncertainty in stress, we calculate the variance across the predicted stress curves from all the ensemble models as $\sigma_S(w, E) = \frac{1}{N-1}\sum^N_i\big(M_i(w, E)-\mu_S(w, E)\big)^2$.

For each structure, we analogously calculate the mean energy dissipation score $D$, our main outcome of interest, by quantifying the area enclosed between the loading and unloading portions of the stress-strain curve using the trapezoidal rule, which we denote as $f_d$. The average and uncertainty in energy dissipation are then assessed via the mean, $\mu_D(w, E) = \frac{1}{N}\sum^N_if_d(M_i(w, E))$, of this energy dissipation area across all predictions and the corresponding variance, $\sigma_D(w, E) = \frac{1}{N-1}\sum^N_i\big(f_d(M_i(w, E))-\mu_D(w, E)\big)^2$ .

\subsection{Energy-dissipation Optimization}

\paragraph{Overview} Our optimization process works as follows. We first acquire a batch of uniformly selected structures in design space, and 3D print and test these structures to get our first training dataset. Our initial batch is uniformly selected, and does not include the TPMS primitives, as is standard in Bayesian optimization to prevent biases in our exploration. We train our neural network and query it with an \textit{acquisition function} to tell us which structures to 3D-print and test next. We then train the network on all available data once again, and repeat the process, iteratively acquiring a total of ten batches of approximately 25 TPMS structures at a time. Each time we query the network for the next batch, we aim to strike a balance between exploring the TPMS performance space (which is more heavily emphasized in our first three batches) and maximizing energy dissipation (which we focus on in the last three batches). For batch selection, we follow the Bayesian-optimization pipeline~\cite{BOsurvey,gonzalez2016batch}. 

\vspace*{-0.05in}
\paragraph{Batch Selection} We begin by choosing \num{1000000} new metamaterial parameter configurations utilizing the Dirichlet distribution \cite{lin2016dirichlet} with $\alpha = 1$ to ensure all parameters sum to unity and are sampled uniformly across the 8-dimensional hyper-simplex. We then apply the surrogate model to each set of parameters to predict stress-strain curves, energy dissipation scores, and associated uncertainty scores. From these \num{1000000} potential designs, we iteratively select the top 40 structures based on their performance as evaluated by the \textit{acquisition function} (details below) and, of these, we fabricate and test an average of 25 structures per batch (as many as time allows per 3D printing slot).

\vspace*{-0.05in}
\paragraph{Acquisition Function}
Our acquisition function informs our selection of the structures to be printed and tested based on a specified objective: either exploration of design space or exploitation of our energy-dissipation performance goal. We base our acquisition on the Upper Confidence Bound Acquisition Function \cite{srinivas2009gaussian}, whereby each sample is scored according to a weighted sum of the uncertainty scores and the mean objective scores: 
\[
\text{UCB}(\mathbf{\params}) = \mu(\params) + \kappa \sigma(\params),
\] \\[-12pt]
where \( \mu(\params) \) is the predicted mean energy dissipation from our model for parameters \( \params \), \( \sigma(\params) \) is the variance energy dissipation, and \( \kappa \) is a trade-off parameter between exploration and exploitation. 

We introduce a localization penalty to reduce excessive clustering when selecting new sample points. During batch selection, we consider a sphere with radius $r$ around all previously selected points from earlier and current batches. If a new sample falls within this radius, it is disallowed. Departing from the theoretical localization penalty suggested in \cite{gonzalez2016batch}, which in our case would give $r\approx0.01$, we use a larger radius of $r=0.2$, as our limited number of larger batches requires a stronger deterrent against clustering.

We evaluate $\text{UCB}(\mathbf{\params})$ for all $\mathbf{\params}$ out of our 1,000,000 samples, iteratively selecting structures for our batch by choosing the one with the highest $UCB$ score that is at least $r=0.2$ away from all other selected structures in previous and current batches. This process continues until we acquire 40 points to 3D print and test in priority order. We repeat this process for all the batches of data.

To balance exploration and exploitation, we adjust the exploration parameter $\kappa$ across batches, starting at $\kappa=2$ for batches 2 and 3 to encourage exploration, progressively reducing it to 1, 0.75, 0.5, and finally 0 for the fourth to tenth batches to focus on optimizing energy dissipation. As shown in \cref{fig:kappa}, high $\kappa$ values favor uncertainty, while low $\kappa$ values favor energy dissipation.

\Cref{fig:network} shows batch-wise experimental and predicted stress-strain curves, where predictions are made by a model trained on all data up to but not including the current batch. As the process moves from exploration to exploitation, the uncertainty in the selected samples decreases, following the typical behavior of batch Bayesian optimization. This shift is also reflected in the improved predictive accuracy with each successive batch, as seen in the figure.

\begin{figure*} 
 \centering
 \includegraphics[width=1.02\textwidth]{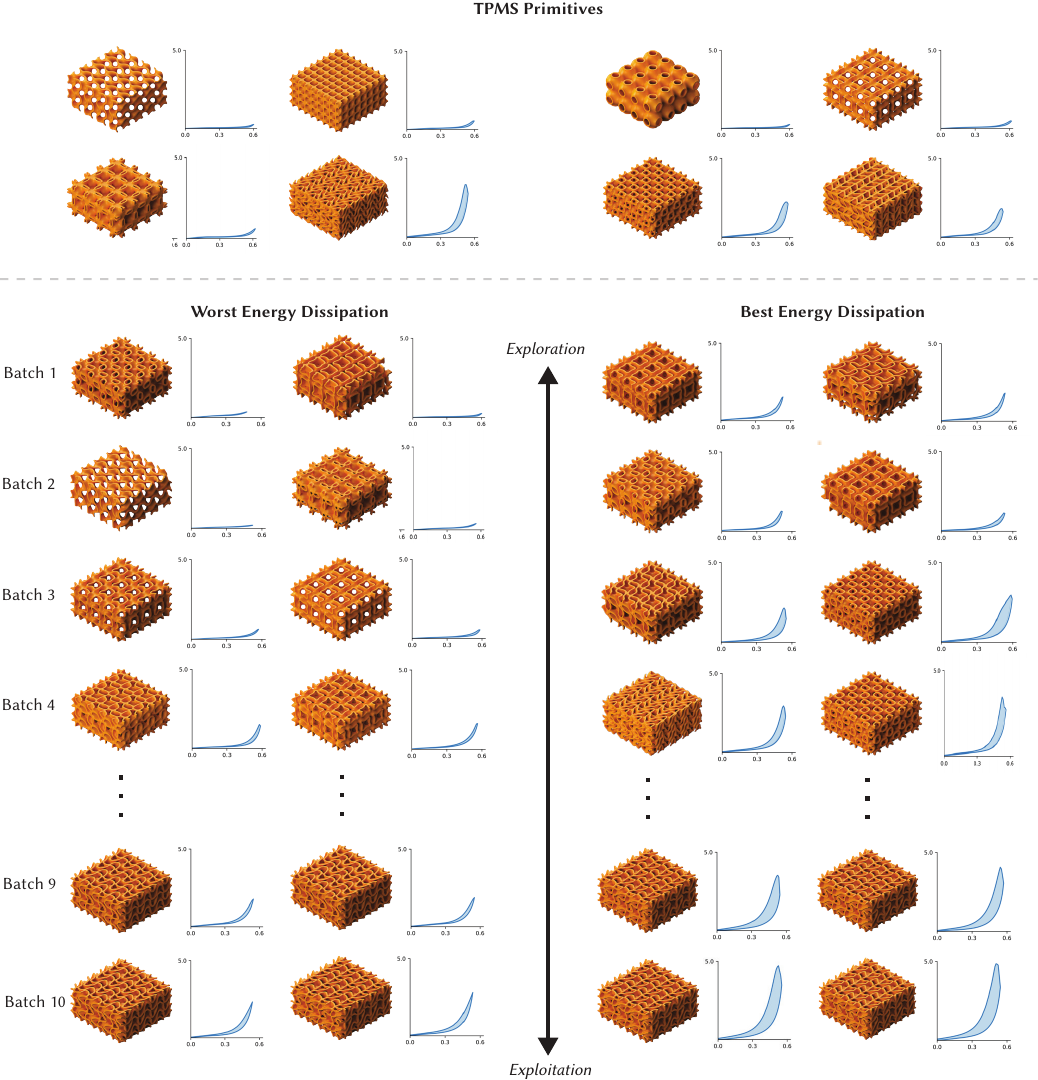}
 \caption{We plot the experimental strain ($x$-axis) and stress ($y$-axis, \unit{\mega\Pa}) curves for the TPMS primitive structures (top) and our batched data (bottom). We display the two samples for each batch with the worst (left two columns) and best (right two columns) energy dissipation (area between loading and unloading stress-strain curves, \unit{k\J/\m^3}, highlighted in light blue). We can see the effects of our optimization procedure on the batched data. As we transition from exploration in batch 2 towards exploitation in batch 10, both the best and worst performing structures of each batch improve, and we are ultimately able to find structures with energy dissipation significantly higher than the initial uniform samples and the TPMS primitives. The highest-energy dissipation structure is that at the bottom right of the plot, with energy dissipation of 435.98\unit{k\J/\m^3}.}
 \Description{}
 \label{fig:batches}
\end{figure*}

\begin{figure}
  \centering
  \vspace*{0.1in}
  \includegraphics[width=0.965\linewidth]{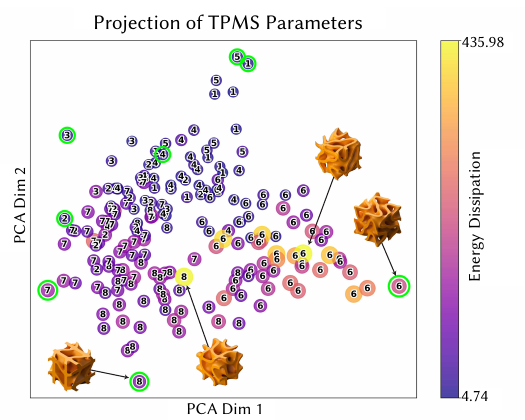}
  \vspace*{-0.1in}
  \caption{\Ac{PCA} projection of metamaterial parameters, with color indicating energy dissipation and labels showing the highest-percentage \ac{TPMS} primitive for each structure. Green circles highlight the single \ac{TPMS} primitives. Structures with the highest percentages of primitives 6 and 8 exhibits the highest energy dissipation, surpassing the performance of those primitives on their own. We highlight a few of our highest-performing mixture structures and their corresponding highest percentage TPMS primitives.}
  \Description{}
  \label{fig:pca}
  \vspace*{-0.07in}
\end{figure}

\vspace*{-0.01in}
\section{Material Discovery Results}

\paragraph{Material Discovery} Our optimization effectively discovers novel TPMS structures with significantly enhanced energy dissipation. \cref{fig:batches} shows the best and worst performing structures from each batch, with experimental stress-strain curves for loading and unloading phases. Our goal is to maximize the area between these curves to optimize energy dissipation. Both the best and worst structures improve throughout the optimization.

\Cref{tab:scores} reports the minimum, maximum, median, and mean energy dissipation of each batch. The initial uniformly sampled batch had a mean energy dissipation of \qty{18.81}{k\J\per\m^3}, while the final optimized batch achieved a mean of \qty{232.39}{k\J\per\m^3}. Our approach identifies structures with energy dissipation twenty times higher than the average for randomly selected structures.

We identified 70 structures with energy dissipation exceeding \qty{165.96}{k\J\per\m^3}, outperforming the best primitive, Fischer-Koch. These materials often resemble Fischer-Koch but include significant proportions of other primitives, enhancing their performance.

\begin{table}
\centering
\small
\vspace*{-0.07in}
\caption{Energy Dissipation Values (\unit{k\J\per\m^3}) by Batch}
\vspace*{-0.12in}
\begin{tabular}{@{}ccccc@{}}
\toprule
\textbf{Batch} & \textbf{Minimum} & \textbf{Maximum} & \textbf{Mean} & \textbf{Median} \\ \midrule
primitives (\textbf{baseline}) &  6.80 & 165.96 & 54.13 & 15.51 \\
1 (initial samples) & 5.81 & 45.77 & 18.81 & 17.92 \\
2 & 4.74 & 44.89 & 15.60 & 12.63 \\
3 & 13.14 & 218.99 & 63.70 & 41.25 \\
4 & 16.48 & 175.21 & 82.64 & 76.12 \\
5 & 24.88 & 196.87 & 81.65 & 77.59 \\
6 & 49.78 & 312.94 & 138.91 & 123.7 \\
7 & 54.96 & 339.18 & 182.30 & 179.69 \\
8 & 104.51 & 226.45 & 168.76 & 170.36 \\
9 & 56.29 & 320.47 & 151.88 & 132.99 \\
10 & 92.74 & 435.98 & 232.39 & 239.55 \\
\bottomrule
\vspace*{-0.21in}
\label{tab:scores}
\end{tabular}

\end{table}

\paragraph{Dataset Analysis}

Our final hyperelastic TPMS dataset includes ten batches and TPMS primitives, labeled with stress-strain curves. Energy dissipation ranges from \qty{4.74}{k\J\per\m^3} to \qty{435.98}{k\J\per\m^3}, over double that of the energy dissipation of Fischer-Koch, which has an energy density of \noindent \qty{165.96}{k\J\per\m^3}. 

We use Principal Component Analysis (PCA) \cite{abdi2010principal} to visualize the dataset's performance, as shown in Figure~\ref{fig:pca}. Each circle's color and size represent energy dissipation, with numerical labels indicating the predominant primitive. Structures 6 and 8 account for most high energy-dissipation instances, while primitives 1, 4, and 5 generally have lower energy dissipation.

\paragraph{Applications} TPMS structures have a wide range of potential applications due to their unique properties. They are lightweight, porous, flexible, and, as we have shown, can be designed to have high energy dissipation. We illustrate three potential applications of our discovered hyperelastic TPMS structures in \cref{fig:applications}: a knee pad, a bone filling, and a robot vacuum cleaner bumper. 

\section{Conclusion}

\paragraph{Limitations and Future Work}
Our method has limitations worth exploring. While effective with TPMS structures, it could be extended to other metamaterials and objectives. Future work should develop a multi-class database of validated hyperelastic metamaterials. Using 4×4×2 lattices may miss behaviors like aperiodic buckling in larger, nonuniform lattices. Additionally, the IP-PDMS photoresist used here, while hyperelastic, dissipates energy less effectively than metals or ceramics because it lacks deformation mechanisms such as plasticity and fracture. Testing optimized TPMS lattices made from stiffer hyperelastic materials could validate applicability for impact mitigation, focusing on material recoverability, peak stress reduction, and energy dissipation. Lastly, while we optimized our model using efficient microscale fabrication techniques and experimental testing methods, to replicate our findings in a performance setting, future work should explore larger length scales and also identify the physical basis of the dissipation mechanisms. 

\paragraph{Summary}
In ten 3-hour 3D-printing sessions, our Bayesian optimization framework effectively discovered  microscale TPMS structures that dissipate over twice the energy of the best-performing Fischer-Koch TPMS. We present the first physically validated microscale hyperelastic TPMS dataset, including 274 structures, 70 of which exhibit properties superior to those of the best primitive structure.

\begin{acks}

We would like to thank Stefanie Mueller for her support throughout this work, Ane Zu\~niga for her valuable insights, and Pavle Konaković for his help with modeling and rendering. We are also grateful to the MIT Morningside Academy for Design, the Andrew (1956) and Erna Viterbi Fellowship, and the Mathworks Fellowship for their support. Financial support from the DEVCOM Army Research Laboratory Army Research Office through the Massachusetts Institute of Technology (MIT) Institute for Soldier Nanotechnologies (ISN) under Cooperative Agreement Number W911NF-23-2-0121 is gratefully acknowledged. This work was carried out in part through the use of the MIT.nano facilities.

\end{acks}

\bibliographystyle{ACM-Reference-Format}
\bibliography{bibliography}


\begin{thebibliography}{81}


\ifx \showCODEN    \undefined \def \showCODEN     #1{\unskip}     \fi
\ifx \showDOI      \undefined \def \showDOI       #1{#1}\fi
\ifx \showISBNx    \undefined \def \showISBNx     #1{\unskip}     \fi
\ifx \showISBNxiii \undefined \def \showISBNxiii  #1{\unskip}     \fi
\ifx \showISSN     \undefined \def \showISSN      #1{\unskip}     \fi
\ifx \showLCCN     \undefined \def \showLCCN      #1{\unskip}     \fi
\ifx \shownote     \undefined \def \shownote      #1{#1}          \fi
\ifx \showarticletitle \undefined \def \showarticletitle #1{#1}   \fi
\ifx \showURL      \undefined \def \showURL       {\relax}        \fi
\providecommand\bibfield[2]{#2}
\providecommand\bibinfo[2]{#2}
\providecommand\natexlab[1]{#1}
\providecommand\showeprint[2][]{arXiv:#2}

\bibitem[Abdi and Williams(2010)]%
        {abdi2010principal}
\bibfield{author}{\bibinfo{person}{Herv{\'e} Abdi} {and}
  \bibinfo{person}{Lynne~J Williams}.} \bibinfo{year}{2010}\natexlab{}.
\newblock \showarticletitle{Principal component analysis}.
\newblock \bibinfo{journal}{\emph{Wiley Interdisciplinary Reviews:
  Computational Statistics}} \bibinfo{volume}{2}, \bibinfo{number}{4}
  (\bibinfo{year}{2010}), \bibinfo{pages}{433--459}.
\newblock


\bibitem[Al-Ketan and Abu Al-Rub(2021)]%
        {AlKetan2021MSLattice}
\bibfield{author}{\bibinfo{person}{Oraib Al-Ketan} {and}
  \bibinfo{person}{Rashid~K. Abu Al-Rub}.} \bibinfo{year}{2021}\natexlab{}.
\newblock \showarticletitle{MSLattice: A free software for generating uniform
  and graded lattices based on triply periodic minimal surfaces}.
\newblock \bibinfo{journal}{\emph{Material Design \& Processing
  Communications}} \bibinfo{volume}{3}, \bibinfo{number}{6}
  (\bibinfo{year}{2021}), \bibinfo{pages}{e205}.
\newblock
\newblock
\shownote{e205 MDPC-2020-042.R1}.


\bibitem[Ashby(2006)]%
        {Ashby2006}
\bibfield{author}{\bibinfo{person}{M.F Ashby}.}
  \bibinfo{year}{2006}\natexlab{}.
\newblock \showarticletitle{The properties of foams and lattices}.
\newblock \bibinfo{journal}{\emph{Philosophical Transactions of the Royal
  Society A: Mathematical, Physical and Engineering Sciences}}
  \bibinfo{volume}{364}, \bibinfo{number}{1838} (\bibinfo{date}{Jan.}
  \bibinfo{year}{2006}), \bibinfo{pages}{15--30}.
\newblock
\showISSN{1364-503X, 1471-2962}
\urldef\tempurl%
\url{https://doi.org/10.1098/rsta.2005.1678}
\showDOI{\tempurl}


\bibitem[Attarzadeh et~al\mbox{.}(2022)]%
        {attarzadeh2022multi}
\bibfield{author}{\bibinfo{person}{Reza Attarzadeh},
  \bibinfo{person}{Seyed-Hosein Attarzadeh-Niaki}, {and}
  \bibinfo{person}{Christophe Duwig}.} \bibinfo{year}{2022}\natexlab{}.
\newblock \showarticletitle{Multi-objective optimization of TPMS-based heat
  exchangers for low-temperature waste heat recovery}.
\newblock \bibinfo{journal}{\emph{Applied Thermal Engineering}}
  \bibinfo{volume}{212} (\bibinfo{year}{2022}), \bibinfo{pages}{118448}.
\newblock


\bibitem[Babaee et~al\mbox{.}(2013)]%
        {Babaee2013}
\bibfield{author}{\bibinfo{person}{Sahab Babaee}, \bibinfo{person}{Jongmin
  Shim}, \bibinfo{person}{James~C. Weaver}, \bibinfo{person}{Elizabeth~R.
  Chen}, \bibinfo{person}{Nikita Patel}, {and} \bibinfo{person}{Katia
  Bertoldi}.} \bibinfo{year}{2013}\natexlab{}.
\newblock \showarticletitle{{3D} {Soft} {Metamaterials} with {Negative}
  {Poisson}'s {Ratio}}.
\newblock \bibinfo{journal}{\emph{Advanced Materials}} \bibinfo{volume}{25},
  \bibinfo{number}{36} (\bibinfo{date}{Sept.} \bibinfo{year}{2013}),
  \bibinfo{pages}{5044--5049}.
\newblock
\showISSN{09359648}
\urldef\tempurl%
\url{https://doi.org/10.1002/adma.201301986}
\showDOI{\tempurl}


\bibitem[Barnes et~al\mbox{.}(2014)]%
        {Barnes2014}
\bibfield{author}{\bibinfo{person}{A.T. Barnes}, \bibinfo{person}{K.
  Ravi-Chandar}, \bibinfo{person}{S. Kyriakides}, {and} \bibinfo{person}{S.
  Gaitanaros}.} \bibinfo{year}{2014}\natexlab{}.
\newblock \showarticletitle{Dynamic crushing of aluminum foams: {Part} {I} –
  {Experiments}}.
\newblock \bibinfo{journal}{\emph{International Journal of Solids and
  Structures}} \bibinfo{volume}{51}, \bibinfo{number}{9} (\bibinfo{date}{May}
  \bibinfo{year}{2014}), \bibinfo{pages}{1631--1645}.
\newblock
\showISSN{00207683}
\urldef\tempurl%
\url{https://doi.org/10.1016/j.ijsolstr.2013.11.019}
\showDOI{\tempurl}


\bibitem[Bastek et~al\mbox{.}(2022)]%
        {bastek2022inverting}
\bibfield{author}{\bibinfo{person}{Jan-Hendrik Bastek},
  \bibinfo{person}{Siddhant Kumar}, \bibinfo{person}{Bastian Telgen},
  \bibinfo{person}{Rapha{\"e}l~N Glaesener}, {and} \bibinfo{person}{Dennis~M
  Kochmann}.} \bibinfo{year}{2022}\natexlab{}.
\newblock \showarticletitle{Inverting the structure--property map of truss
  metamaterials by deep learning}.
\newblock \bibinfo{journal}{\emph{Proceedings of the National Academy of
  Sciences}} \bibinfo{volume}{119}, \bibinfo{number}{1} (\bibinfo{year}{2022}),
  \bibinfo{pages}{e2111505119}.
\newblock


\bibitem[Bauer et~al\mbox{.}(2017)]%
        {Bauer2017}
\bibfield{author}{\bibinfo{person}{Jens Bauer}, \bibinfo{person}{Lucas~R.
  Meza}, \bibinfo{person}{Tobias~A. Schaedler}, \bibinfo{person}{Ruth
  Schwaiger}, \bibinfo{person}{Xiaoyu Zheng}, {and} \bibinfo{person}{Lorenzo
  Valdevit}.} \bibinfo{year}{2017}\natexlab{}.
\newblock \showarticletitle{Nanolattices: An emerging class of mechanical
  metamaterials}.
\newblock \bibinfo{journal}{\emph{Advanced Materials}} \bibinfo{volume}{29},
  \bibinfo{number}{40} (\bibinfo{year}{2017}), \bibinfo{pages}{1701850}.
\newblock
\urldef\tempurl%
\url{https://doi.org/10.1002/adma.201701850}
\showDOI{\tempurl}
\showeprint{https://onlinelibrary.wiley.com/doi/pdf/10.1002/adma.201701850}


\bibitem[Bickel et~al\mbox{.}(2010)]%
        {Bickel2010}
\bibfield{author}{\bibinfo{person}{Bernd Bickel}, \bibinfo{person}{Moritz
  Bächer}, \bibinfo{person}{Miguel~A. Otaduy}, \bibinfo{person}{Hyunho~Richard
  Lee}, \bibinfo{person}{Hanspeter Pfister}, \bibinfo{person}{Markus Gross},
  {and} \bibinfo{person}{Wojciech Matusik}.} \bibinfo{year}{2010}\natexlab{}.
\newblock \showarticletitle{Design and Fabrication of Materials with Desired
  Deformation Behavior}. In \bibinfo{booktitle}{\emph{ACM SIGGRAPH 2010
  Papers}}. \bibinfo{publisher}{Association for Computing Machinery},
  \bibinfo{address}{New York, NY, USA}, \bibinfo{pages}{1--10}.
\newblock
\urldef\tempurl%
\url{https://doi.org/10.1145/1833349.1778800}
\showDOI{\tempurl}


\bibitem[Caruana et~al\mbox{.}(2000)]%
        {caruana2000overfitting}
\bibfield{author}{\bibinfo{person}{Rich Caruana}, \bibinfo{person}{Steve
  Lawrence}, {and} \bibinfo{person}{C Giles}.} \bibinfo{year}{2000}\natexlab{}.
\newblock \showarticletitle{Overfitting in neural nets: Backpropagation,
  conjugate gradient, and early stopping}.
\newblock \bibinfo{journal}{\emph{Advances in Neural Information Processing
  Systems}}  \bibinfo{volume}{13} (\bibinfo{year}{2000}),
  \bibinfo{pages}{381--387}.
\newblock


\bibitem[Chen et~al\mbox{.}(2021)]%
        {Chen2021Bistable}
\bibfield{author}{\bibinfo{person}{Tian Chen}, \bibinfo{person}{Julian
  Panetta}, \bibinfo{person}{Max Schnaubelt}, {and} \bibinfo{person}{Mark
  Pauly}.} \bibinfo{year}{2021}\natexlab{}.
\newblock \showarticletitle{Bistable auxetic surface structures}.
\newblock \bibinfo{journal}{\emph{ACM Trans. Graph.}} \bibinfo{volume}{40},
  \bibinfo{number}{4}, Article \bibinfo{articleno}{39} (\bibinfo{date}{jul}
  \bibinfo{year}{2021}), \bibinfo{numpages}{9}~pages.
\newblock
\showISSN{0730-0301}


\bibitem[Crook et~al\mbox{.}(2020)]%
        {Crook2020}
\bibfield{author}{\bibinfo{person}{Cameron Crook}, \bibinfo{person}{Jens
  Bauer}, \bibinfo{person}{Anna Guell~Izard}, \bibinfo{person}{Cristine
  Santos~de Oliveira}, \bibinfo{person}{Juliana Martins de Souza~e Silva},
  \bibinfo{person}{Jonathan~B. Berger}, {and} \bibinfo{person}{Lorenzo
  Valdevit}.} \bibinfo{year}{2020}\natexlab{}.
\newblock \showarticletitle{Plate-nanolattices at the theoretical limit of
  stiffness and strength}.
\newblock \bibinfo{journal}{\emph{Nature Communications}} \bibinfo{volume}{11},
  \bibinfo{number}{1} (\bibinfo{year}{2020}), \bibinfo{pages}{1579}.
\newblock
\showISBNx{2041-1723}
\urldef\tempurl%
\url{https://doi.org/10.1038/s41467-020-15434-2}
\showDOI{\tempurl}


\bibitem[Dattelbaum et~al\mbox{.}(2020)]%
        {Dattelbaum2020}
\bibfield{author}{\bibinfo{person}{Dana~M. Dattelbaum}, \bibinfo{person}{Axinte
  Ionita}, \bibinfo{person}{Brian~M. Patterson}, \bibinfo{person}{Brittany~A.
  Branch}, {and} \bibinfo{person}{Lindsey Kuettner}.}
  \bibinfo{year}{2020}\natexlab{}.
\newblock \showarticletitle{Shockwave dissipation by interface-dominated porous
  structures}.
\newblock \bibinfo{journal}{\emph{AIP Advances}} \bibinfo{volume}{10},
  \bibinfo{number}{7} (\bibinfo{date}{July} \bibinfo{year}{2020}),
  \bibinfo{pages}{075016}.
\newblock
\showISSN{2158-3226}
\urldef\tempurl%
\url{https://doi.org/10.1063/5.0015179}
\showDOI{\tempurl}


\bibitem[Deshpande et~al\mbox{.}(2001)]%
        {Deshpande2001}
\bibfield{author}{\bibinfo{person}{V.S. Deshpande}, \bibinfo{person}{M.F.
  Ashby}, {and} \bibinfo{person}{N.A. Fleck}.} \bibinfo{year}{2001}\natexlab{}.
\newblock \showarticletitle{Foam topology: Bending versus stretching dominated
  architectures}.
\newblock \bibinfo{journal}{\emph{Acta Materialia}} \bibinfo{volume}{49},
  \bibinfo{number}{6} (\bibinfo{date}{April} \bibinfo{year}{2001}),
  \bibinfo{pages}{1035--1040}.
\newblock
\showISSN{13596454}
\urldef\tempurl%
\url{https://doi.org/10.1016/S1359-6454(00)00379-7}
\showDOI{\tempurl}


\bibitem[Deshpande and Fleck(2000)]%
        {Deshpande2000}
\bibfield{author}{\bibinfo{person}{V.S. Deshpande} {and} \bibinfo{person}{N.A.
  Fleck}.} \bibinfo{year}{2000}\natexlab{}.
\newblock \showarticletitle{High strain rate compressive behaviour of aluminium
  alloy foams}.
\newblock \bibinfo{journal}{\emph{International Journal of Impact Engineering}}
  \bibinfo{volume}{24}, \bibinfo{number}{3} (\bibinfo{date}{March}
  \bibinfo{year}{2000}), \bibinfo{pages}{277--298}.
\newblock
\showISSN{0734743X}
\urldef\tempurl%
\url{https://doi.org/10.1016/S0734-743X(99)00153-0}
\showDOI{\tempurl}


\bibitem[Dong and Zhao(2021)]%
        {dong2021application}
\bibfield{author}{\bibinfo{person}{Zhifei Dong} {and} \bibinfo{person}{Xin
  Zhao}.} \bibinfo{year}{2021}\natexlab{}.
\newblock \showarticletitle{Application of TPMS structure in bone
  regeneration}.
\newblock \bibinfo{journal}{\emph{Engineered Regeneration}}
  \bibinfo{volume}{2} (\bibinfo{year}{2021}), \bibinfo{pages}{154--162}.
\newblock


\bibitem[Erps et~al\mbox{.}(2021)]%
        {Erps2021}
\bibfield{author}{\bibinfo{person}{Timothy Erps}, \bibinfo{person}{Michael
  Foshey}, \bibinfo{person}{Mina~Konaković Luković}, \bibinfo{person}{Wan
  Shou}, \bibinfo{person}{Hanns~Hagen Goetzke}, \bibinfo{person}{Herve
  Dietsch}, \bibinfo{person}{Klaus Stoll}, \bibinfo{person}{Bernhard von
  Vacano}, {and} \bibinfo{person}{Wojciech Matusik}.}
  \bibinfo{year}{2021}\natexlab{}.
\newblock \showarticletitle{Accelerated discovery of 3D printing materials
  using data-driven multiobjective optimization}.
\newblock \bibinfo{journal}{\emph{Science Advances}} \bibinfo{volume}{7},
  \bibinfo{number}{42} (\bibinfo{year}{2021}), \bibinfo{pages}{eabf7435}.
\newblock
\urldef\tempurl%
\url{https://doi.org/10.1126/sciadv.abf7435}
\showDOI{\tempurl}
\showeprint{https://www.science.org/doi/pdf/10.1126/sciadv.abf7435}


\bibitem[Fan et~al\mbox{.}(2022)]%
        {fan2022thermal}
\bibfield{author}{\bibinfo{person}{Zhaohui Fan}, \bibinfo{person}{Renjing Gao},
  {and} \bibinfo{person}{Shutian Liu}.} \bibinfo{year}{2022}\natexlab{}.
\newblock \showarticletitle{Thermal conductivity enhancement and thermal
  saturation elimination designs of battery thermal management system for phase
  change materials based on triply periodic minimal surface}.
\newblock \bibinfo{journal}{\emph{Energy}}  \bibinfo{volume}{259}
  (\bibinfo{year}{2022}), \bibinfo{pages}{125091}.
\newblock


\bibitem[Fang and Zhan(2019)]%
        {fang2019deep}
\bibfield{author}{\bibinfo{person}{Zhiwei Fang} {and} \bibinfo{person}{Justin
  Zhan}.} \bibinfo{year}{2019}\natexlab{}.
\newblock \showarticletitle{Deep physical informed neural networks for
  metamaterial design}.
\newblock \bibinfo{journal}{\emph{IEEE Access}}  \bibinfo{volume}{8}
  (\bibinfo{year}{2019}), \bibinfo{pages}{24506--24513}.
\newblock


\bibitem[Feng et~al\mbox{.}(2021)]%
        {feng2021isotropic}
\bibfield{author}{\bibinfo{person}{Jiawei Feng}, \bibinfo{person}{Bo Liu},
  \bibinfo{person}{Zhiwei Lin}, {and} \bibinfo{person}{Jianzhong Fu}.}
  \bibinfo{year}{2021}\natexlab{}.
\newblock \showarticletitle{Isotropic porous structure design methods based on
  triply periodic minimal surfaces}.
\newblock \bibinfo{journal}{\emph{Materials \& Design}}  \bibinfo{volume}{210}
  (\bibinfo{year}{2021}), \bibinfo{pages}{110050}.
\newblock


\bibitem[Gatt and Grima(2008)]%
        {gatt2008negative}
\bibfield{author}{\bibinfo{person}{Ruben Gatt} {and} \bibinfo{person}{Joseph~N
  Grima}.} \bibinfo{year}{2008}\natexlab{}.
\newblock \showarticletitle{Negative compressibility}.
\newblock \bibinfo{journal}{\emph{physica status solidi (RRL)--Rapid Research
  Letters}} \bibinfo{volume}{2}, \bibinfo{number}{5} (\bibinfo{year}{2008}),
  \bibinfo{pages}{236--238}.
\newblock


\bibitem[Gonz{\'a}lez et~al\mbox{.}(2016)]%
        {gonzalez2016batch}
\bibfield{author}{\bibinfo{person}{Javier Gonz{\'a}lez},
  \bibinfo{person}{Zhenwen Dai}, \bibinfo{person}{Philipp Hennig}, {and}
  \bibinfo{person}{Neil Lawrence}.} \bibinfo{year}{2016}\natexlab{}.
\newblock \showarticletitle{Batch Bayesian optimization via local
  penalization}. In \bibinfo{booktitle}{\emph{Artificial Intelligence and
  Statistics}}. PMLR, \bibinfo{pages}{648--657}.
\newblock


\bibitem[Ha et~al\mbox{.}(2023)]%
        {ha2023rapid}
\bibfield{author}{\bibinfo{person}{Chan~Soo Ha}, \bibinfo{person}{Desheng Yao},
  \bibinfo{person}{Zhenpeng Xu}, \bibinfo{person}{Chenang Liu},
  \bibinfo{person}{Han Liu}, \bibinfo{person}{Daniel Elkins},
  \bibinfo{person}{Matthew Kile}, \bibinfo{person}{Vikram Deshpande},
  \bibinfo{person}{Zhenyu Kong}, \bibinfo{person}{Mathieu Bauchy},
  {et~al\mbox{.}}} \bibinfo{year}{2023}\natexlab{}.
\newblock \showarticletitle{Rapid inverse design of metamaterials based on
  prescribed mechanical behavior through machine learning}.
\newblock \bibinfo{journal}{\emph{Nature Communications}} \bibinfo{volume}{14},
  \bibinfo{number}{1} (\bibinfo{year}{2023}), \bibinfo{pages}{5765}.
\newblock


\bibitem[Hawreliak et~al\mbox{.}(2016)]%
        {Hawreliak2016}
\bibfield{author}{\bibinfo{person}{J.~A. Hawreliak}, \bibinfo{person}{J. Lind},
  \bibinfo{person}{B. Maddox}, \bibinfo{person}{M. Barham}, \bibinfo{person}{M.
  Messner}, \bibinfo{person}{N. Barton}, \bibinfo{person}{B.~J. Jensen}, {and}
  \bibinfo{person}{M. Kumar}.} \bibinfo{year}{2016}\natexlab{}.
\newblock \showarticletitle{Dynamic {Behavior} of {Engineered} {Lattice}
  {Materials}}.
\newblock \bibinfo{journal}{\emph{Scientific Reports}} \bibinfo{volume}{6},
  \bibinfo{number}{1} (\bibinfo{date}{June} \bibinfo{year}{2016}),
  \bibinfo{pages}{28094}.
\newblock
\showISSN{2045-2322}
\urldef\tempurl%
\url{https://doi.org/10.1038/srep28094}
\showDOI{\tempurl}


\bibitem[Hu et~al\mbox{.}(2020)]%
        {hu2020efficient}
\bibfield{author}{\bibinfo{person}{Jiangbei Hu}, \bibinfo{person}{Shengfa
  Wang}, \bibinfo{person}{Baojun Li}, \bibinfo{person}{Fengqi Li},
  \bibinfo{person}{Zhongxuan Luo}, {and} \bibinfo{person}{Ligang Liu}.}
  \bibinfo{year}{2020}\natexlab{}.
\newblock \showarticletitle{Efficient representation and optimization for
  TPMS-based porous structures}.
\newblock \bibinfo{journal}{\emph{IEEE Transactions on Visualization and
  Computer Graphics}} \bibinfo{volume}{28}, \bibinfo{number}{7}
  (\bibinfo{year}{2020}), \bibinfo{pages}{2615--2627}.
\newblock


\bibitem[Huang et~al\mbox{.}(2024)]%
        {Huang2024Differentiable}
\bibfield{author}{\bibinfo{person}{Zizhou Huang}, \bibinfo{person}{Davi~Colli
  Tozoni}, \bibinfo{person}{Arvi Gjoka}, \bibinfo{person}{Zachary Ferguson},
  \bibinfo{person}{Teseo Schneider}, \bibinfo{person}{Daniele Panozzo}, {and}
  \bibinfo{person}{Denis Zorin}.} \bibinfo{year}{2024}\natexlab{}.
\newblock \showarticletitle{Differentiable solver for time-dependent
  deformation problems with contact}.
\newblock \bibinfo{journal}{\emph{ACM Transactions on Graphics}}
  \bibinfo{volume}{43}, \bibinfo{number}{3} (\bibinfo{date}{May}
  \bibinfo{year}{2024}), \bibinfo{pages}{1--30}.
\newblock
\showISSN{0730-0301}
\urldef\tempurl%
\url{https://doi.org/10.1145/3657648}
\showDOI{\tempurl}


\bibitem[Jiao et~al\mbox{.}(2023)]%
        {jiao2023mechanical}
\bibfield{author}{\bibinfo{person}{Pengcheng Jiao}, \bibinfo{person}{Jochen
  Mueller}, \bibinfo{person}{Jordan~R Raney}, \bibinfo{person}{Xiaoyu Zheng},
  {and} \bibinfo{person}{Amir~H Alavi}.} \bibinfo{year}{2023}\natexlab{}.
\newblock \showarticletitle{Mechanical metamaterials and beyond}.
\newblock \bibinfo{journal}{\emph{Nature Communications}} \bibinfo{volume}{14},
  \bibinfo{number}{1} (\bibinfo{year}{2023}), \bibinfo{pages}{6004}.
\newblock


\bibitem[Jones et~al\mbox{.}(1998)]%
        {jones1998efficient}
\bibfield{author}{\bibinfo{person}{Donald~R Jones}, \bibinfo{person}{Matthias
  Schonlau}, {and} \bibinfo{person}{William~J Welch}.}
  \bibinfo{year}{1998}\natexlab{}.
\newblock \showarticletitle{Efficient global optimization of expensive
  black-box functions}.
\newblock \bibinfo{journal}{\emph{Journal of Global Optimization}}
  \bibinfo{volume}{13}, \bibinfo{number}{4} (\bibinfo{year}{1998}),
  \bibinfo{pages}{455}.
\newblock


\bibitem[Kuszczak et~al\mbox{.}(2023)]%
        {honeycomb2023}
\bibfield{author}{\bibinfo{person}{I. Kuszczak}, \bibinfo{person}{F.I. Azam},
  \bibinfo{person}{M.A. Bessa}, \bibinfo{person}{P.J. Tan}, {and}
  \bibinfo{person}{F. Bosi}.} \bibinfo{year}{2023}\natexlab{}.
\newblock \showarticletitle{Bayesian optimisation of hexagonal honeycomb
  metamaterial}.
\newblock \bibinfo{journal}{\emph{Extreme Mechanics Letters}}
  \bibinfo{volume}{64} (\bibinfo{year}{2023}), \bibinfo{pages}{102078}.
\newblock
\showISSN{2352-4316}
\urldef\tempurl%
\url{https://doi.org/10.1016/j.eml.2023.102078}
\showDOI{\tempurl}


\bibitem[Lakshminarayanan et~al\mbox{.}(2017)]%
        {DE2017}
\bibfield{author}{\bibinfo{person}{Balaji Lakshminarayanan},
  \bibinfo{person}{Alexander Pritzel}, {and} \bibinfo{person}{Charles
  Blundell}.} \bibinfo{year}{2017}\natexlab{}.
\newblock \showarticletitle{Simple and S scalable predictive incertainty
  estimation using deep ensembles}. In \bibinfo{booktitle}{\emph{Advances in
  Neural Information Processing Systems 30 (NeurIPS 2017)}}.
  \bibinfo{publisher}{Curran Associates, Inc.}, \bibinfo{pages}{6402--6413}.
\newblock
\urldef\tempurl%
\url{https://proceedings.neurips.cc/paper_files/paper/2017/file/9ef2ed4b7fd2c810847ffa5fa85bce38-Paper.pdf}
\showURL{%
\tempurl}


\bibitem[Lee et~al\mbox{.}(2024)]%
        {huang2023optimized}
\bibfield{author}{\bibinfo{person}{Chan-Lock~A. Lee}, \bibinfo{person}{Zizhou
  Huang}, \bibinfo{person}{Daniele Panozzo}, {and} \bibinfo{person}{Denis
  Zorin}.} \bibinfo{year}{2024}\natexlab{}.
\newblock \showarticletitle{Computational design of flexible planar
  microstructures}.
\newblock \bibinfo{journal}{\emph{ACM Transactions on Graphics}}
  \bibinfo{volume}{43}, \bibinfo{number}{6} (\bibinfo{year}{2024}),
  \bibinfo{pages}{1--21}.
\newblock
\urldef\tempurl%
\url{https://doi.org/10.1145/3687765}
\showDOI{\tempurl}


\bibitem[Lee et~al\mbox{.}(2023)]%
        {lee2022t}
\bibfield{author}{\bibinfo{person}{Doksoo Lee}, \bibinfo{person}{Yu-Chin Chan},
  \bibinfo{person}{Wei Chen}, \bibinfo{person}{Liwei Wang},
  \bibinfo{person}{Anton van Beek}, {and} \bibinfo{person}{Wei Chen}.}
  \bibinfo{year}{2023}\natexlab{}.
\newblock \showarticletitle{t-METASET: Task-aware acquisition of metamaterial
  datasets through diversity-based active learning}.
\newblock \bibinfo{journal}{\emph{Journal of Mechanical Design}}
  \bibinfo{volume}{145}, \bibinfo{number}{3} (\bibinfo{date}{March}
  \bibinfo{year}{2023}), \bibinfo{pages}{031704}.
\newblock
\showISSN{1050-0472}
\urldef\tempurl%
\url{https://doi.org/10.1115/1.4055925}
\showDOI{\tempurl}


\bibitem[Li et~al\mbox{.}(2022)]%
        {Li2022Digital}
\bibfield{author}{\bibinfo{person}{Weichen Li}, \bibinfo{person}{Fengwen Wang},
  \bibinfo{person}{Ole Sigmund}, {and} \bibinfo{person}{Xiaojia~Shelly Zhang}.}
  \bibinfo{year}{2022}\natexlab{}.
\newblock \showarticletitle{Digital synthesis of free-form multimaterial
  structures for realization of arbitrary programmed mechanical responses}.
\newblock \bibinfo{journal}{\emph{Proceedings of the National Academy of
  Sciences}} \bibinfo{volume}{119}, \bibinfo{number}{10}
  (\bibinfo{year}{2022}), \bibinfo{pages}{e2120563119}.
\newblock
\urldef\tempurl%
\url{https://doi.org/10.1073/pnas.2120563119}
\showDOI{\tempurl}
\showeprint{https://www.pnas.org/doi/pdf/10.1073/pnas.2120563119}


\bibitem[Li et~al\mbox{.}(2023)]%
        {li2023neural}
\bibfield{author}{\bibinfo{person}{Yue Li}, \bibinfo{person}{Stelian Coros},
  {and} \bibinfo{person}{Bernhard Thomaszewski}.}
  \bibinfo{year}{2023}\natexlab{}.
\newblock \showarticletitle{Neural metamaterial networks for nonlinear material
  design}.
\newblock \bibinfo{journal}{\emph{ACM Transactions on Graphics (TOG)}}
  \bibinfo{volume}{42}, \bibinfo{number}{6} (\bibinfo{year}{2023}),
  \bibinfo{pages}{1--13}.
\newblock


\bibitem[Lim et~al\mbox{.}(2021)]%
        {lim2021extrapolative}
\bibfield{author}{\bibinfo{person}{Yee-Fun Lim}, \bibinfo{person}{Chee~Koon
  Ng}, \bibinfo{person}{US Vaitesswar}, {and} \bibinfo{person}{Kedar
  Hippalgaonkar}.} \bibinfo{year}{2021}\natexlab{}.
\newblock \showarticletitle{Extrapolative Bayesian optimization with Gaussian
  process and neural network ensemble surrogate models}.
\newblock \bibinfo{journal}{\emph{Advanced Intelligent Systems}}
  \bibinfo{volume}{3}, \bibinfo{number}{11} (\bibinfo{year}{2021}),
  \bibinfo{pages}{2100101}.
\newblock


\bibitem[Lin(2016)]%
        {lin2016dirichlet}
\bibfield{author}{\bibinfo{person}{Jiayu Lin}.}
  \bibinfo{year}{2016}\natexlab{}.
\newblock \showarticletitle{On the {D}irichlet distribution}.
\newblock \bibinfo{journal}{\emph{Department of Mathematics and Statistics,
  Queens University}}  \bibinfo{volume}{40} (\bibinfo{year}{2016}).
\newblock


\bibitem[Lind et~al\mbox{.}(2019)]%
        {Lind2019}
\bibfield{author}{\bibinfo{person}{Jonathan Lind}, \bibinfo{person}{Brian~J.
  Jensen}, \bibinfo{person}{Matthew Barham}, {and} \bibinfo{person}{Mukul
  Kumar}.} \bibinfo{year}{2019}\natexlab{}.
\newblock \showarticletitle{In situ dynamic compression wave behavior in
  additively manufactured lattice materials}.
\newblock \bibinfo{journal}{\emph{Journal of Materials Research}}
  \bibinfo{volume}{34}, \bibinfo{number}{1} (\bibinfo{date}{Jan.}
  \bibinfo{year}{2019}), \bibinfo{pages}{2--19}.
\newblock
\showISSN{0884-2914, 2044-5326}
\urldef\tempurl%
\url{https://doi.org/10.1557/jmr.2018.351}
\showDOI{\tempurl}


\bibitem[Liu et~al\mbox{.}(2024)]%
        {liu2024programmable}
\bibfield{author}{\bibinfo{person}{Chenyang Liu}, \bibinfo{person}{Xi Zhang},
  \bibinfo{person}{Jiahui Chang}, \bibinfo{person}{You Lyu},
  \bibinfo{person}{Jianan Zhao}, {and} \bibinfo{person}{Song Qiu}.}
  \bibinfo{year}{2024}\natexlab{}.
\newblock \showarticletitle{Programmable mechanical metamaterials: Basic
  concepts, types, construction strategies—a review}.
\newblock \bibinfo{journal}{\emph{Frontiers in Materials}}
  \bibinfo{volume}{11} (\bibinfo{year}{2024}), \bibinfo{pages}{1361408}.
\newblock


\bibitem[Makatura et~al\mbox{.}(2023)]%
        {makatura2023procedural}
\bibfield{author}{\bibinfo{person}{Liane Makatura}, \bibinfo{person}{Bohan
  Wang}, \bibinfo{person}{Yi-Lu Chen}, \bibinfo{person}{Bolei Deng},
  \bibinfo{person}{Chris Wojtan}, \bibinfo{person}{Bernd Bickel}, {and}
  \bibinfo{person}{Wojciech Matusik}.} \bibinfo{year}{2023}\natexlab{}.
\newblock \showarticletitle{Procedural metamaterials: A unified procedural
  graph for metamaterial design}.
\newblock \bibinfo{journal}{\emph{ACM Transactions on Graphics}}
  \bibinfo{volume}{42}, \bibinfo{number}{5} (\bibinfo{year}{2023}),
  \bibinfo{pages}{1--19}.
\newblock


\bibitem[Mart\'{\i}nez et~al\mbox{.}(2016)]%
        {Martinez2016}
\bibfield{author}{\bibinfo{person}{Jon\`{a}s Mart\'{\i}nez},
  \bibinfo{person}{J\'{e}r\'{e}mie Dumas}, {and} \bibinfo{person}{Sylvain
  Lefebvre}.} \bibinfo{year}{2016}\natexlab{}.
\newblock \showarticletitle{Procedural Voronoi foams for additive
  manufacturing}.
\newblock \bibinfo{journal}{\emph{ACM Trans. Graph.}} \bibinfo{volume}{35},
  \bibinfo{number}{4}, Article \bibinfo{articleno}{44} (\bibinfo{date}{jul}
  \bibinfo{year}{2016}), \bibinfo{numpages}{12}~pages.
\newblock
\showISSN{0730-0301}
\urldef\tempurl%
\url{https://doi.org/10.1145/2897824.2925922}
\showDOI{\tempurl}


\bibitem[Mart\'{\i}nez et~al\mbox{.}(2018)]%
        {Martinez2018}
\bibfield{author}{\bibinfo{person}{Jon\`{a}s Mart\'{\i}nez},
  \bibinfo{person}{Samuel Hornus}, \bibinfo{person}{Haichuan Song}, {and}
  \bibinfo{person}{Sylvain Lefebvre}.} \bibinfo{year}{2018}\natexlab{}.
\newblock \showarticletitle{Polyhedral Voronoi diagrams for additive
  manufacturing}.
\newblock \bibinfo{journal}{\emph{ACM Trans. Graph.}} \bibinfo{volume}{37},
  \bibinfo{number}{4}, Article \bibinfo{articleno}{129} (\bibinfo{date}{jul}
  \bibinfo{year}{2018}), \bibinfo{numpages}{15}~pages.
\newblock
\showISSN{0730-0301}
\urldef\tempurl%
\url{https://doi.org/10.1145/3197517.3201343}
\showDOI{\tempurl}


\bibitem[Matlack et~al\mbox{.}(2016)]%
        {Matlack2016}
\bibfield{author}{\bibinfo{person}{Kathryn~H. Matlack}, \bibinfo{person}{Anton
  Bauhofer}, \bibinfo{person}{Sebastian Krödel}, \bibinfo{person}{Antonio
  Palermo}, {and} \bibinfo{person}{Chiara Daraio}.}
  \bibinfo{year}{2016}\natexlab{}.
\newblock \showarticletitle{Composite 3D-printed metastructures for
  low-frequency and broadband vibration absorption}.
\newblock \bibinfo{journal}{\emph{Proceedings of the National Academy of
  Sciences}} \bibinfo{volume}{113}, \bibinfo{number}{30}
  (\bibinfo{year}{2016}), \bibinfo{pages}{8386--8390}.
\newblock
\urldef\tempurl%
\url{https://doi.org/10.1073/pnas.1600171113}
\showDOI{\tempurl}
\showeprint{https://www.pnas.org/doi/pdf/10.1073/pnas.1600171113}


\bibitem[Meyer et~al\mbox{.}(2022)]%
        {meyer2022graph}
\bibfield{author}{\bibinfo{person}{Paul~P Meyer}, \bibinfo{person}{Colin
  Bonatti}, \bibinfo{person}{Thomas Tancogne-Dejean}, {and}
  \bibinfo{person}{Dirk Mohr}.} \bibinfo{year}{2022}\natexlab{}.
\newblock \showarticletitle{Graph-based metamaterials: Deep learning of
  structure-property relations}.
\newblock \bibinfo{journal}{\emph{Materials \& Design}}  \bibinfo{volume}{223}
  (\bibinfo{year}{2022}), \bibinfo{pages}{111175}.
\newblock


\bibitem[Mines et~al\mbox{.}(2013)]%
        {Mines2013}
\bibfield{author}{\bibinfo{person}{R.A.W. Mines}, \bibinfo{person}{S.
  Tsopanos}, \bibinfo{person}{Y. Shen}, \bibinfo{person}{R. Hasan}, {and}
  \bibinfo{person}{S.T. McKown}.} \bibinfo{year}{2013}\natexlab{}.
\newblock \showarticletitle{Drop weight impact behaviour of sandwich panels
  with metallic micro lattice cores}.
\newblock \bibinfo{journal}{\emph{International Journal of Impact Engineering}}
   \bibinfo{volume}{60} (\bibinfo{date}{Oct.} \bibinfo{year}{2013}),
  \bibinfo{pages}{120--132}.
\newblock
\showISSN{0734743X}
\urldef\tempurl%
\url{https://doi.org/10.1016/j.ijimpeng.2013.04.007}
\showDOI{\tempurl}


\bibitem[Nakshatrala et~al\mbox{.}(2013)]%
        {Nakshatrala2013Nonlinear}
\bibfield{author}{\bibinfo{person}{Praveen~Babu Nakshatrala},
  \bibinfo{person}{Daniel~A Tortorelli}, {and} \bibinfo{person}{KB3069875
  Nakshatrala}.} \bibinfo{year}{2013}\natexlab{}.
\newblock \showarticletitle{Nonlinear structural design using multiscale
  topology optimization. Part I: Static formulation}.
\newblock \bibinfo{journal}{\emph{Computer Methods in Applied Mechanics and
  Engineering}}  \bibinfo{volume}{261} (\bibinfo{year}{2013}),
  \bibinfo{pages}{167--176}.
\newblock


\bibitem[Nicolaou and Motter(2012)]%
        {nicolaou2012mechanical}
\bibfield{author}{\bibinfo{person}{Zachary~G Nicolaou} {and}
  \bibinfo{person}{Adilson~E Motter}.} \bibinfo{year}{2012}\natexlab{}.
\newblock \showarticletitle{Mechanical metamaterials with negative
  compressibility transitions}.
\newblock \bibinfo{journal}{\emph{Nature Materials}} \bibinfo{volume}{11},
  \bibinfo{number}{7} (\bibinfo{year}{2012}), \bibinfo{pages}{608--613}.
\newblock


\bibitem[Novak et~al\mbox{.}(2023)]%
        {Novak2023}
\bibfield{author}{\bibinfo{person}{Nejc Novak}, \bibinfo{person}{Oraib
  Al-Ketan}, \bibinfo{person}{Anja Mauko}, \bibinfo{person}{Yunus~Emre Yilmaz},
  \bibinfo{person}{Lovre Krstulović-Opara}, \bibinfo{person}{Shigeru Tanaka},
  \bibinfo{person}{Kazuyuki Hokamoto}, \bibinfo{person}{Reza Rowshan},
  \bibinfo{person}{Rashid~Abu Al-Rub}, \bibinfo{person}{Matej Vesenjak}, {and}
  \bibinfo{person}{Zoran Ren}.} \bibinfo{year}{2023}\natexlab{}.
\newblock \showarticletitle{Impact loading of additively manufactured metallic
  stochastic sheet-based cellular material}.
\newblock \bibinfo{journal}{\emph{International Journal of Impact Engineering}}
   \bibinfo{volume}{174} (\bibinfo{date}{April} \bibinfo{year}{2023}),
  \bibinfo{pages}{104527}.
\newblock
\showISSN{0734743X}
\urldef\tempurl%
\url{https://doi.org/10.1016/j.ijimpeng.2023.104527}
\showDOI{\tempurl}


\bibitem[Oktay et~al\mbox{.}(2023)]%
        {oktay2023neuromechanical}
\bibfield{author}{\bibinfo{person}{Deniz Oktay}, \bibinfo{person}{Mehran
  Mirramezani}, \bibinfo{person}{Eder Medina}, {and} \bibinfo{person}{Ryan~P
  Adams}.} \bibinfo{year}{2023}\natexlab{}.
\newblock \showarticletitle{Neuromechanical autoencoders: Learning to couple
  elastic and neural network nonlinearity}. In \bibinfo{booktitle}{\emph{The
  Eleventh International Conference on Learning Representations}}.
\newblock
\urldef\tempurl%
\url{https://openreview.net/forum?id=QubsmJT_A0}
\showURL{%
\tempurl}


\bibitem[Panetta et~al\mbox{.}(2017)]%
        {panetta2017worst}
\bibfield{author}{\bibinfo{person}{Julian Panetta}, \bibinfo{person}{Abtin
  Rahimian}, {and} \bibinfo{person}{Denis Zorin}.}
  \bibinfo{year}{2017}\natexlab{}.
\newblock \showarticletitle{Worst-case stress relief for microstructures}.
\newblock \bibinfo{journal}{\emph{ACM Transactions on Graphics (TOG)}}
  \bibinfo{volume}{36}, \bibinfo{number}{4} (\bibinfo{year}{2017}),
  \bibinfo{pages}{1--16}.
\newblock


\bibitem[Panetta et~al\mbox{.}(2015)]%
        {panetta2015elastic}
\bibfield{author}{\bibinfo{person}{Julian Panetta}, \bibinfo{person}{Qingnan
  Zhou}, \bibinfo{person}{Luigi Malomo}, \bibinfo{person}{Nico Pietroni},
  \bibinfo{person}{Paolo Cignoni}, {and} \bibinfo{person}{Denis Zorin}.}
  \bibinfo{year}{2015}\natexlab{}.
\newblock \showarticletitle{Elastic textures for additive fabrication}.
\newblock \bibinfo{journal}{\emph{ACM Transactions on Graphics (TOG)}}
  \bibinfo{volume}{34}, \bibinfo{number}{4} (\bibinfo{year}{2015}),
  \bibinfo{pages}{1--12}.
\newblock


\bibitem[Peng et~al\mbox{.}(2023)]%
        {peng2023machine}
\bibfield{author}{\bibinfo{person}{Bo Peng}, \bibinfo{person}{Ye Wei},
  \bibinfo{person}{Yu Qin}, \bibinfo{person}{Jiabao Dai}, \bibinfo{person}{Yue
  Li}, \bibinfo{person}{Aobo Liu}, \bibinfo{person}{Yun Tian},
  \bibinfo{person}{Liuliu Han}, \bibinfo{person}{Yufeng Zheng}, {and}
  \bibinfo{person}{Peng Wen}.} \bibinfo{year}{2023}\natexlab{}.
\newblock \showarticletitle{Machine learning-enabled constrained
  multi-objective design of architected materials}.
\newblock \bibinfo{journal}{\emph{Nature Communications}} \bibinfo{volume}{14},
  \bibinfo{number}{1} (\bibinfo{year}{2023}), \bibinfo{pages}{6630}.
\newblock


\bibitem[Santurkar et~al\mbox{.}(2018)]%
        {santurkar2018does}
\bibfield{author}{\bibinfo{person}{Shibani Santurkar},
  \bibinfo{person}{Dimitris Tsipras}, \bibinfo{person}{Andrew Ilyas}, {and}
  \bibinfo{person}{Aleksander Madry}.} \bibinfo{year}{2018}\natexlab{}.
\newblock \showarticletitle{How does batch normalization help optimization?}.
  In \bibinfo{booktitle}{\emph{Advances in Neural Information Processing
  Systems 31 (NeurIPS 2018)}}. \bibinfo{publisher}{Curran Associates, Inc.},
  \bibinfo{pages}{2488--2498}.
\newblock
\urldef\tempurl%
\url{https://proceedings.neurips.cc/paper_files/paper/2018/file/90556c1c1a5d7c2c4f5f7f1e1c1e1e1e-Paper.pdf}
\showURL{%
\tempurl}


\bibitem[Schaedler et~al\mbox{.}(2011)]%
        {schaedler2011ultralight}
\bibfield{author}{\bibinfo{person}{Tobias~A Schaedler}, \bibinfo{person}{Alan~J
  Jacobsen}, \bibinfo{person}{Anna Torrents}, \bibinfo{person}{Adam~E
  Sorensen}, \bibinfo{person}{Jie Lian}, \bibinfo{person}{Julia~R Greer},
  \bibinfo{person}{Lorenzo Valdevit}, {and} \bibinfo{person}{Wiliam~B Carter}.}
  \bibinfo{year}{2011}\natexlab{}.
\newblock \showarticletitle{Ultralight metallic microlattices}.
\newblock \bibinfo{journal}{\emph{Science}} \bibinfo{volume}{334},
  \bibinfo{number}{6058} (\bibinfo{year}{2011}), \bibinfo{pages}{962--965}.
\newblock


\bibitem[Schafer(2011)]%
        {schafer2011savitzky}
\bibfield{author}{\bibinfo{person}{Ronald~W Schafer}.}
  \bibinfo{year}{2011}\natexlab{}.
\newblock \showarticletitle{What is a Savitzky-Golay filter? [Lecture notes]}.
\newblock \bibinfo{journal}{\emph{IEEE Signal Processing Magazine}}
  \bibinfo{volume}{28}, \bibinfo{number}{4} (\bibinfo{year}{2011}),
  \bibinfo{pages}{111--117}.
\newblock


\bibitem[Schneider et~al\mbox{.}(2019)]%
        {polyfem}
\bibfield{author}{\bibinfo{person}{Teseo Schneider}, \bibinfo{person}{Jérémie
  Dumas}, \bibinfo{person}{Xifeng Gao}, \bibinfo{person}{Denis Zorin}, {and}
  \bibinfo{person}{Daniele Panozzo}.} \bibinfo{year}{2019}\natexlab{}.
\newblock \bibinfo{title}{{PolyFEM}}.
\newblock \bibinfo{howpublished}{\url{https://polyfem.github.io/}}.
\newblock


\bibitem[Schumacher et~al\mbox{.}(2015)]%
        {Schumacher2015Microstructures}
\bibfield{author}{\bibinfo{person}{Christian Schumacher},
  \bibinfo{person}{Bernd Bickel}, \bibinfo{person}{Jan Rys},
  \bibinfo{person}{Steve Marschner}, \bibinfo{person}{Chiara Daraio}, {and}
  \bibinfo{person}{Markus Gross}.} \bibinfo{year}{2015}\natexlab{}.
\newblock \showarticletitle{Microstructures to control elasticity in 3D
  printing}.
\newblock \bibinfo{journal}{\emph{ACM Trans. Graph.}} \bibinfo{volume}{34},
  \bibinfo{number}{4}, Article \bibinfo{articleno}{136} (\bibinfo{date}{jul}
  \bibinfo{year}{2015}), \bibinfo{numpages}{13}~pages.
\newblock
\showISSN{0730-0301}
\urldef\tempurl%
\url{https://doi.org/10.1145/2766926}
\showDOI{\tempurl}


\bibitem[Serles et~al\mbox{.}(2025)]%
        {serles2025ultrahigh}
\bibfield{author}{\bibinfo{person}{Peter Serles}, \bibinfo{person}{Jinwook
  Yeo}, \bibinfo{person}{Michel Hach{\'e}}, \bibinfo{person}{Pedro~Guerra
  Demingos}, \bibinfo{person}{Jonathan Kong}, \bibinfo{person}{Pascal Kiefer},
  \bibinfo{person}{Somayajulu Dhulipala}, \bibinfo{person}{Boran Kumral},
  \bibinfo{person}{Katherine Jia}, \bibinfo{person}{Shuo Yang},
  {et~al\mbox{.}}} \bibinfo{year}{2025}\natexlab{}.
\newblock \showarticletitle{Ultrahigh specific strength by Bayesian
  optimization of carbon nanolattices}.
\newblock \bibinfo{journal}{\emph{Advanced Materials}} \bibinfo{volume}{37},
  \bibinfo{number}{14} (\bibinfo{year}{2025}), \bibinfo{pages}{2570108}.
\newblock


\bibitem[{Shahriari} et~al\mbox{.}(2016)]%
        {BOsurvey}
\bibfield{author}{\bibinfo{person}{B. {Shahriari}}, \bibinfo{person}{K.
  {Swersky}}, \bibinfo{person}{Z. {Wang}}, \bibinfo{person}{R.~P. {Adams}},
  {and} \bibinfo{person}{N. {de Freitas}}.} \bibinfo{year}{2016}\natexlab{}.
\newblock \showarticletitle{Taking the human out of the loop: A review of
  Bayesian Optimization}.
\newblock \bibinfo{journal}{\emph{Proc. IEEE}} \bibinfo{volume}{104},
  \bibinfo{number}{1} (\bibinfo{year}{2016}), \bibinfo{pages}{148--175}.
\newblock


\bibitem[Shaikeea et~al\mbox{.}(2022)]%
        {shaikeea2022}
\bibfield{author}{\bibinfo{person}{Angkur Jyoti~Dipanka Shaikeea},
  \bibinfo{person}{Huachen Cui}, \bibinfo{person}{Mark O’Masta},
  \bibinfo{person}{Xiaoyu~Rayne Zheng}, {and} \bibinfo{person}{Vikram~Sudhir
  Deshpande}.} \bibinfo{year}{2022}\natexlab{}.
\newblock \showarticletitle{The toughness of mechanical metamaterials}.
\newblock \bibinfo{journal}{\emph{Nature Materials}} \bibinfo{volume}{21},
  \bibinfo{number}{3} (\bibinfo{date}{March} \bibinfo{year}{2022}),
  \bibinfo{pages}{297--304}.
\newblock
\showISSN{1476-1122, 1476-4660}
\urldef\tempurl%
\url{https://doi.org/10.1038/s41563-021-01182-1}
\showDOI{\tempurl}


\bibitem[Sharpe et~al\mbox{.}(2018)]%
        {Sharpe2018}
\bibfield{author}{\bibinfo{person}{C. Sharpe}, \bibinfo{person}{C. Seepersad},
  \bibinfo{person}{D.~A. Tortorelli}, {and} \bibinfo{person}{S.~E. Watts}.}
  \bibinfo{year}{2018}\natexlab{}.
\newblock \showarticletitle{Design of mechanical metamaterials via constrained
  Bayesian optimization}. In \bibinfo{booktitle}{\emph{Proceedings of the ASME
  2018 International Design Engineering Technical Conferences and Computers and
  Information in Engineering Conference (IDETC/CIE 2018)}}.
\newblock
\urldef\tempurl%
\url{https://doi.org/10.1115/DETC2018-85270}
\showDOI{\tempurl}


\bibitem[Shim et~al\mbox{.}(2013)]%
        {Shim2013}
\bibfield{author}{\bibinfo{person}{Jongmin Shim}, \bibinfo{person}{Sicong
  Shan}, \bibinfo{person}{Andrej Ko{\v s}mrlj}, \bibinfo{person}{Sung~H. Kang},
  \bibinfo{person}{Elizabeth~R. Chen}, \bibinfo{person}{James~C. Weaver}, {and}
  \bibinfo{person}{Katia Bertoldi}.} \bibinfo{year}{2013}\natexlab{}.
\newblock \showarticletitle{Harnessing instabilities for design of soft
  reconfigurable auxetic/chiral materials}.
\newblock \bibinfo{journal}{\emph{Soft Matter}}  \bibinfo{volume}{9}
  (\bibinfo{year}{2013}), \bibinfo{pages}{8198--8202}.
\newblock
Issue 34.
\urldef\tempurl%
\url{https://doi.org/10.1039/C3SM51148K}
\showDOI{\tempurl}


\bibitem[Snapp et~al\mbox{.}(2024)]%
        {snapp2024superlative}
\bibfield{author}{\bibinfo{person}{Kelsey~L Snapp}, \bibinfo{person}{Benjamin
  Verdier}, \bibinfo{person}{Aldair~E Gongora}, \bibinfo{person}{Samuel
  Silverman}, \bibinfo{person}{Adedire~D Adesiji}, \bibinfo{person}{Elise~F
  Morgan}, \bibinfo{person}{Timothy~J Lawton}, \bibinfo{person}{Emily Whiting},
  {and} \bibinfo{person}{Keith~A Brown}.} \bibinfo{year}{2024}\natexlab{}.
\newblock \showarticletitle{Superlative mechanical energy absorbing efficiency
  discovered through self-driving lab-human partnership}.
\newblock \bibinfo{journal}{\emph{Nature Communications}} \bibinfo{volume}{15},
  \bibinfo{number}{1} (\bibinfo{year}{2024}), \bibinfo{pages}{4290}.
\newblock


\bibitem[Srinivas et~al\mbox{.}(2010)]%
        {srinivas2009gaussian}
\bibfield{author}{\bibinfo{person}{Niranjan Srinivas}, \bibinfo{person}{Andreas
  Krause}, \bibinfo{person}{Sham Kakade}, {and} \bibinfo{person}{Matthias~W.
  Seeger}.} \bibinfo{year}{2010}\natexlab{}.
\newblock \showarticletitle{Gaussian process optimization in the bandit
  setting: {N}o regret and experimental design}. In
  \bibinfo{booktitle}{\emph{Proceedings of the 27th International Conference on
  Machine Learning (ICML)}}. \bibinfo{publisher}{Omnipress},
  \bibinfo{pages}{1015--1022}.
\newblock
\urldef\tempurl%
\url{http://www.icml2010.org/papers/422.pdf}
\showURL{%
\tempurl}


\bibitem[Srivastava et~al\mbox{.}(2014)]%
        {srivastava2014dropout}
\bibfield{author}{\bibinfo{person}{Nitish Srivastava},
  \bibinfo{person}{Geoffrey Hinton}, \bibinfo{person}{Alex Krizhevsky},
  \bibinfo{person}{Ilya Sutskever}, {and} \bibinfo{person}{Ruslan
  Salakhutdinov}.} \bibinfo{year}{2014}\natexlab{}.
\newblock \showarticletitle{Dropout: A simple way to prevent neural networks
  from overfitting}.
\newblock \bibinfo{journal}{\emph{Journal of Machine Learning Research}}
  \bibinfo{volume}{15}, \bibinfo{number}{1} (\bibinfo{year}{2014}),
  \bibinfo{pages}{1929--1958}.
\newblock


\bibitem[Surjadi and Portela(2025)]%
        {Surjadi2024}
\bibfield{author}{\bibinfo{person}{James~Utama Surjadi} {and}
  \bibinfo{person}{Carlos~M. Portela}.} \bibinfo{year}{2025}\natexlab{}.
\newblock \showarticletitle{Enabling three-dimensional architected materials
  across length scales and timescales}.
\newblock \bibinfo{journal}{\emph{Nature Materials}} \bibinfo{volume}{24},
  \bibinfo{number}{4} (\bibinfo{date}{April} \bibinfo{year}{2025}),
  \bibinfo{pages}{493--505}.
\newblock
\urldef\tempurl%
\url{https://doi.org/10.1038/s41563-025-02119-8}
\showDOI{\tempurl}


\bibitem[Tancogne-Dejean et~al\mbox{.}(2019)]%
        {Tancogne-Dejean2019}
\bibfield{author}{\bibinfo{person}{T. Tancogne-Dejean}, \bibinfo{person}{X.
  Li}, \bibinfo{person}{M. Diamantopoulou}, \bibinfo{person}{C.~C. Roth}, {and}
  \bibinfo{person}{D. Mohr}.} \bibinfo{year}{2019}\natexlab{}.
\newblock \showarticletitle{High strain rate response of
  {additively}-{manufactured} {plate}-{lattices}: {E}xperiments and
  {modeling}}.
\newblock \bibinfo{journal}{\emph{Journal of Dynamic Behavior of Materials}}
  \bibinfo{volume}{5}, \bibinfo{number}{3} (\bibinfo{date}{Sept.}
  \bibinfo{year}{2019}), \bibinfo{pages}{361--375}.
\newblock
\showISSN{2199-7446, 2199-7454}
\urldef\tempurl%
\url{https://doi.org/10.1007/s40870-019-00219-6}
\showDOI{\tempurl}


\bibitem[Thakolkaran et~al\mbox{.}(2025)]%
        {thakolkaran2023experiment}
\bibfield{author}{\bibinfo{person}{Prakash Thakolkaran},
  \bibinfo{person}{Michael Espinal}, \bibinfo{person}{Somayajulu Dhulipala},
  \bibinfo{person}{Siddhant Kumar}, {and} \bibinfo{person}{Carlos~M. Portela}.}
  \bibinfo{year}{2025}\natexlab{}.
\newblock \showarticletitle{Experiment-informed finite-strain inverse design of
  spinodal metamaterials}.
\newblock \bibinfo{journal}{\emph{Extreme Mechanics Letters}}
  \bibinfo{volume}{74} (\bibinfo{date}{Jan.} \bibinfo{year}{2025}),
  \bibinfo{pages}{102274}.
\newblock
\showISSN{23524316}
\urldef\tempurl%
\url{https://doi.org/10.1016/j.eml.2024.102274}
\showDOI{\tempurl}


\bibitem[{The CGAL Project}(2024)]%
        {cgal}
\bibfield{author}{\bibinfo{person}{{The CGAL Project}}.}
  \bibinfo{year}{2024}\natexlab{}.
\newblock \bibinfo{booktitle}{\emph{{CGAL}} (\bibinfo{edition}{{5.6.1}} ed.)}.
\newblock
\urldef\tempurl%
\url{https://www.cgal.org/}
\showURL{%
\tempurl}


\bibitem[Tian et~al\mbox{.}(2024)]%
        {tian2024boundary}
\bibfield{author}{\bibinfo{person}{Yunsheng Tian}, \bibinfo{person}{Ane
  Zuniga}, \bibinfo{person}{Xinwei Zhang}, \bibinfo{person}{Johannes~P.
  Dürholt}, \bibinfo{person}{Payel Das}, \bibinfo{person}{Jie Chen},
  \bibinfo{person}{Wojciech Matusik}, {and} \bibinfo{person}{Mina
  Konaković~Luković}.} \bibinfo{year}{2024}\natexlab{}.
\newblock \showarticletitle{Boundary exploration for Bayesian optimization with
  unknown physical constraints}. In \bibinfo{booktitle}{\emph{Proceedings of
  the 41st International Conference on Machine Learning (ICML)}}.
  \bibinfo{publisher}{PMLR}, \bibinfo{pages}{48295--48320}.
\newblock
\urldef\tempurl%
\url{https://arxiv.org/abs/2402.07692}
\showURL{%
\tempurl}


\bibitem[Tozoni et~al\mbox{.}(2020)]%
        {Tzoni2020Low}
\bibfield{author}{\bibinfo{person}{Davi~Colli Tozoni},
  \bibinfo{person}{J\'{e}r\'{e}mie Dumas}, \bibinfo{person}{Zhongshi Jiang},
  \bibinfo{person}{Julian Panetta}, \bibinfo{person}{Daniele Panozzo}, {and}
  \bibinfo{person}{Denis Zorin}.} \bibinfo{year}{2020}\natexlab{}.
\newblock \showarticletitle{A low-parametric rhombic microstructure family for
  irregular lattices}.
\newblock \bibinfo{journal}{\emph{ACM Trans. Graph.}} \bibinfo{volume}{39},
  \bibinfo{number}{4}, Article \bibinfo{articleno}{101} (\bibinfo{date}{aug}
  \bibinfo{year}{2020}), \bibinfo{numpages}{20}~pages.
\newblock
\showISSN{0730-0301}


\bibitem[Walser(2001)]%
        {walser2001electromagnetic}
\bibfield{author}{\bibinfo{person}{Rodger~M Walser}.}
  \bibinfo{year}{2001}\natexlab{}.
\newblock \showarticletitle{Electromagnetic metamaterials}. In
  \bibinfo{booktitle}{\emph{Complex Mediums II: Beyond Linear Isotropic
  Dielectrics}}, Vol.~\bibinfo{volume}{4467}. SPIE, \bibinfo{pages}{1--15}.
\newblock


\bibitem[Wang et~al\mbox{.}(2020)]%
        {wang2020deep}
\bibfield{author}{\bibinfo{person}{Liwei Wang}, \bibinfo{person}{Yu-Chin Chan},
  \bibinfo{person}{Faez Ahmed}, \bibinfo{person}{Zhao Liu},
  \bibinfo{person}{Ping Zhu}, {and} \bibinfo{person}{Wei Chen}.}
  \bibinfo{year}{2020}\natexlab{}.
\newblock \showarticletitle{Deep generative modeling for mechanistic-based
  learning and design of metamaterial systems}.
\newblock \bibinfo{journal}{\emph{Computer Methods in Applied Mechanics and
  Engineering}}  \bibinfo{volume}{372} (\bibinfo{year}{2020}),
  \bibinfo{pages}{113377}.
\newblock


\bibitem[Wang et~al\mbox{.}(2022)]%
        {Wang2022}
\bibfield{author}{\bibinfo{person}{Yujia Wang}, \bibinfo{person}{Xuan Zhang},
  \bibinfo{person}{Zihe Li}, \bibinfo{person}{Huajian Gao}, {and}
  \bibinfo{person}{Xiaoyan Li}.} \bibinfo{year}{2022}\natexlab{}.
\newblock \showarticletitle{Achieving the theoretical limit of strength in
  shell-based carbon nanolattices}.
\newblock \bibinfo{journal}{\emph{Proceedings of the National Academy of
  Sciences}} \bibinfo{volume}{119}, \bibinfo{number}{34} (\bibinfo{date}{Aug.}
  \bibinfo{year}{2022}), \bibinfo{pages}{e2119536119}.
\newblock
\showISSN{0027-8424, 1091-6490}
\urldef\tempurl%
\url{https://doi.org/10.1073/pnas.2119536119}
\showDOI{\tempurl}


\bibitem[Weeks and Ravichandran(2023)]%
        {Weeks2023}
\bibfield{author}{\bibinfo{person}{J.~S. Weeks} {and} \bibinfo{person}{G.
  Ravichandran}.} \bibinfo{year}{2023}\natexlab{}.
\newblock \showarticletitle{Effect of {topology} on {transient} {dynamic} and
  {shock} {response} of {polymeric} {lattice} {structures}}.
\newblock \bibinfo{journal}{\emph{Journal of Dynamic Behavior of Materials}}
  \bibinfo{volume}{9}, \bibinfo{number}{1} (\bibinfo{date}{March}
  \bibinfo{year}{2023}), \bibinfo{pages}{44--64}.
\newblock
\showISSN{2199-7446, 2199-7454}
\urldef\tempurl%
\url{https://doi.org/10.1007/s40870-022-00359-2}
\showDOI{\tempurl}


\bibitem[Xia et~al\mbox{.}(2022)]%
        {Xia2022}
\bibfield{author}{\bibinfo{person}{Xiaoxing Xia},
  \bibinfo{person}{Christopher~M. Spadaccini}, {and} \bibinfo{person}{Julia~R.
  Greer}.} \bibinfo{year}{2022}\natexlab{}.
\newblock \showarticletitle{Responsive materials architected in space and
  time}.
\newblock \bibinfo{journal}{\emph{Nature Reviews Materials}}
  \bibinfo{volume}{7}, \bibinfo{number}{9} (\bibinfo{year}{2022}),
  \bibinfo{pages}{683--701}.
\newblock
\showISBNx{2058-8437}
\urldef\tempurl%
\url{https://doi.org/10.1038/s41578-022-00450-z}
\showDOI{\tempurl}


\bibitem[Zadpoor et~al\mbox{.}(2023)]%
        {zadpoor2023design}
\bibfield{author}{\bibinfo{person}{Amir~A. Zadpoor},
  \bibinfo{person}{Mohammad~J. Mirzaali}, \bibinfo{person}{Lorenzo Valdevit},
  {and} \bibinfo{person}{Jonathan~B. Hopkins}.}
  \bibinfo{year}{2023}\natexlab{}.
\newblock \showarticletitle{Design, material, function, and fabrication of
  metamaterials}.
\newblock \bibinfo{journal}{\emph{APL Materials}} \bibinfo{volume}{11},
  \bibinfo{number}{2} (\bibinfo{year}{2023}), \bibinfo{pages}{020401}.
\newblock
\urldef\tempurl%
\url{https://doi.org/10.1063/5.0144454}
\showDOI{\tempurl}


\bibitem[Zhang et~al\mbox{.}(2022)]%
        {zhang2022uncertainty}
\bibfield{author}{\bibinfo{person}{Hengrui Zhang}, \bibinfo{person}{Wei Chen},
  \bibinfo{person}{Akshay Iyer}, \bibinfo{person}{Daniel~W Apley}, {and}
  \bibinfo{person}{Wei Chen}.} \bibinfo{year}{2022}\natexlab{}.
\newblock \showarticletitle{Uncertainty-aware mixed-variable machine learning
  for materials design}.
\newblock \bibinfo{journal}{\emph{Scientific Reports}} \bibinfo{volume}{12},
  \bibinfo{number}{1} (\bibinfo{year}{2022}), \bibinfo{pages}{19760}.
\newblock


\bibitem[Zhang et~al\mbox{.}(2023)]%
        {zhang2023study}
\bibfield{author}{\bibinfo{person}{Jing Zhang}, \bibinfo{person}{Suchao Xie},
  \bibinfo{person}{Tao Li}, \bibinfo{person}{Zinan Liu},
  \bibinfo{person}{Shiwei Zheng}, {and} \bibinfo{person}{Hui Zhou}.}
  \bibinfo{year}{2023}\natexlab{}.
\newblock \showarticletitle{A study of multi-stage energy absorption
  characteristics of hybrid sheet TPMS lattices}.
\newblock \bibinfo{journal}{\emph{Thin-Walled Structures}}
  \bibinfo{volume}{190} (\bibinfo{year}{2023}), \bibinfo{pages}{110989}.
\newblock


\bibitem[Zhang et~al\mbox{.}(2020)]%
        {Zhang2020}
\bibfield{author}{\bibinfo{person}{Xuan Zhang}, \bibinfo{person}{Yujia Wang},
  \bibinfo{person}{Bin Ding}, {and} \bibinfo{person}{Xiaoyan Li}.}
  \bibinfo{year}{2020}\natexlab{}.
\newblock \showarticletitle{Design, fabrication, and mechanics of 3D
  micro-/nanolattices}.
\newblock \bibinfo{journal}{\emph{Small}} \bibinfo{volume}{16},
  \bibinfo{number}{15} (\bibinfo{year}{2020}), \bibinfo{pages}{e1902842}.
\newblock
\urldef\tempurl%
\url{https://doi.org/10.1002/smll.201902842}
\showDOI{\tempurl}


\bibitem[Zhao et~al\mbox{.}(2024)]%
        {zhao2024compressive}
\bibfield{author}{\bibinfo{person}{Xuejin Zhao}, \bibinfo{person}{Zhenzong Li},
  \bibinfo{person}{Yupeng Zou}, {and} \bibinfo{person}{Xiaoyu Zhao}.}
  \bibinfo{year}{2024}\natexlab{}.
\newblock \showarticletitle{Compressive characteristics and energy absorption
  capacity of automobile energy-absorbing box with filled porous TPMS
  structures}.
\newblock \bibinfo{journal}{\emph{Applied Sciences}} \bibinfo{volume}{14},
  \bibinfo{number}{9} (\bibinfo{year}{2024}), \bibinfo{pages}{3790}.
\newblock


\bibitem[Zheng et~al\mbox{.}(2014)]%
        {Zheng2014ultralight}
\bibfield{author}{\bibinfo{person}{Xiaoyu Zheng}, \bibinfo{person}{Howon Lee},
  \bibinfo{person}{Todd~H. Weisgraber}, \bibinfo{person}{Maxim Shusteff},
  \bibinfo{person}{Joshua DeOtte}, \bibinfo{person}{Eric~B. Duoss},
  \bibinfo{person}{Joshua~D. Kuntz}, \bibinfo{person}{Monika~M. Biener},
  \bibinfo{person}{Qi Ge}, \bibinfo{person}{Julie~A. Jackson},
  \bibinfo{person}{Sergei~O. Kucheyev}, \bibinfo{person}{Nicholas~X. Fang},
  {and} \bibinfo{person}{Christopher~M. Spadaccini}.}
  \bibinfo{year}{2014}\natexlab{}.
\newblock \showarticletitle{Ultralight, ultrastiff mechanical metamaterials}.
\newblock \bibinfo{journal}{\emph{Science}} \bibinfo{volume}{344},
  \bibinfo{number}{6190} (\bibinfo{year}{2014}), \bibinfo{pages}{1373--1377}.
\newblock
\urldef\tempurl%
\url{https://doi.org/10.1126/science.1252291}
\showDOI{\tempurl}
\showeprint{https://www.science.org/doi/pdf/10.1126/science.1252291}


\end{thebibliography}

\clearpage

\includepdf[pages=-]{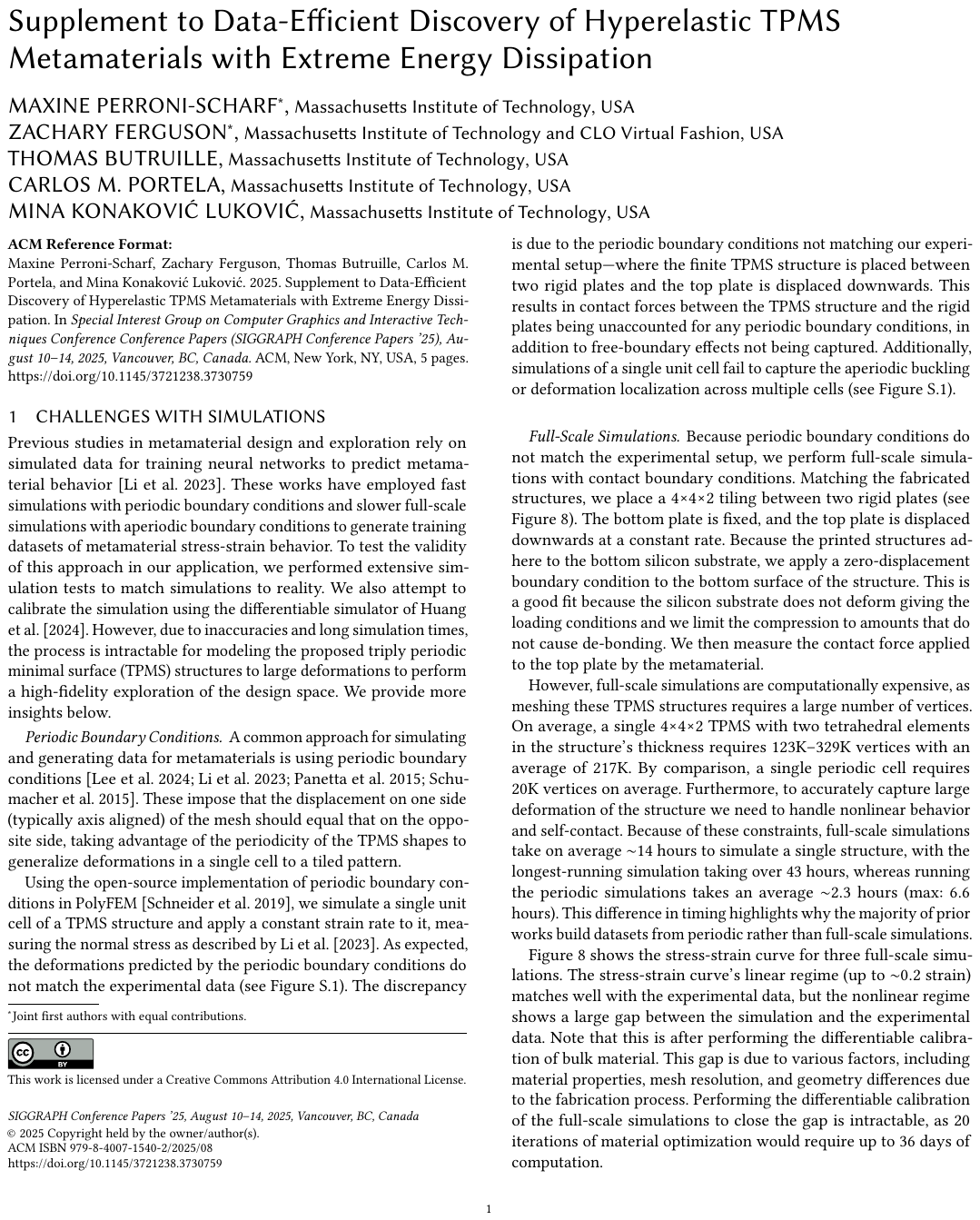}

\end{document}